\documentclass[11pt]{article}
\usepackage{amsmath,amssymb}
\usepackage{amsthm}
\newtheorem{proposition}{Proposition}
\usepackage{latexsym}
\usepackage{psfrag}
\usepackage{epsfig}
\usepackage{graphicx,subfigure}
\usepackage{graphicx}
\usepackage{color, times, verbatim, graphics, array, colordvi}
\usepackage{mathrsfs}
\usepackage{epstopdf}
\usepackage{cite}
\usepackage{caption}
\newtheorem{assumption}{Assumption}

\newtheorem{lemma}{Lemma}
\newtheorem{remark}{Remark}
\newtheorem{theorem}{Theorem}

\newtheorem{problem}{Problem}

\begin{document}
\date{}
\title{Discrete-time SIS Social Contagion Processes on Hypergraphs}
\author{Lidan Liang, Shaoxuan Cui, and Fangzhou Liu
\thanks{L. Liang and F. Liu are with the School of Astronautics, Harbin Institute of Technology, Harbin 150001, China (e-mail: lidan.liang@stu.hit.edu.cn; fangzhou.liu@hit.edu.cn)}
\thanks{Shaoxuan Cui is with the Bernoulli Institute, Faculty of Science and Engineering, University of Groningen, 9700 AB Groningen, The Netherlands (e-mail: s.cui@rug.nl)}}
\maketitle
{\bf Abstract:}
Recent research on social contagion processes has revealed the limitations of traditional networks, which capture only pairwise relationships, to characterize complex multiparty relationships and group influences properly. Social contagion processes on higher-order networks (simplicial complexes and general hypergraphs) have therefore emerged as a novel frontier. In this work, we investigate discrete-time Susceptible-Infected-Susceptible (SIS) social contagion processes occurring on weighted and directed hypergraphs and their extensions to bivirus cases and general higher-order SIS processes with the aid of tensor algebra. Our focus lies in comprehensively characterizing the healthy state and endemic equilibria within this framework. The emergence of bistability or multistability behavior phenomena, where multiple equilibria coexist and are simultaneously locally asymptotically stable, is demonstrated in view of the presence of the higher-order interaction. The novel sufficient conditions of the appearance for system behaviors, which are determined by both (higher-order) network topology and transition rates, are provided to assess the likelihood of the SIS social contagion processes causing an outbreak. More importantly, given the equilibrium is locally stable, an explicit domain of attraction associated with the system parameters is constructed. Moreover, a learning method to estimate the transition rates is presented. In the end, the attained theoretical results are supplemented via numerical examples. Specifically, we evaluate the effectiveness of the networked SIS social contagion process by comparing it with the $2^n$-state Markov chain model. These numerical examples are given to highlight the performance of parameter learning algorithms and the system behaviors of the discrete-time SIS social contagion process.


\section{Introduction}
~~~~Social contagion processes are ubiquitous in daily lives and exert profound and extensive influence. Owing primarily to their powerful applications, including but not limited to marketing and public health, the study of social contagion processes has attracted great attention over the past decades \cite{leskovec2007dynamics,hsiang2020effect,saberi2020simple}. In a broad sense, these processes cover the spread of behaviors, emotions, attitudes, or opinions among individuals within a social network through direct or indirect interactions \cite{ugander2012structural,christakis2013social}. To date, a large number of representative achievements have been reported for the research of social contagion processes regarding their modeling, analysis, and control \cite{zhou2020realistic,radel2010social,liu2020optimal}.

Mathematical modeling of social contagion processes has been a crucial research domain and multitudes of models have been proposed to characterize the underlying mechanism from diverse points of view. Among these models, epidemic models have received considerable attention on account of the potential to describe general diffusion processes, such as information dissemination in social networks and virus propagation \cite{chamley2013models,xia2020dynamic}. As a typical kind of the most popular epidemic models, compartmental models, e.g., susceptible-infected-susceptible (SIS) models and Susceptible-Infected-Recovered (SIR), have a long-standing history that can be traced back to the seminal achievements \cite{sahneh2013generalized,pare2018analysis,liu2019analysis,aleja2022compartmental}. As reported in \cite{sahneh2013generalized}, compartmental models have been thoroughly explored and applied at both micro and macro levels. The Markov chain compartmental models emphasize the micro-level transitions of individual states. To streamline the investigation of epidemic dynamics and manage larger, more complicated system, researchers have incorporated the mean-field compartmental models, building upon the foundation of the Markov chain models. This mean-field model utilizes a macroscopic lens, supplanting intricate individual interactions with average effects, thereby facilitating more operational system analysis. At present, both the continuous and discrete variants of the mean-field models have been formulated and successfully employed in the examination of a multitude of disease system \cite{pare2018analysis,liu2019analysis}. 

On the other hand, the network topology undeniably plays a crucial role in determining the emerging collective behavior for social contagion processes. However, the vast majority of networked social contagion processes have been constructed based on traditional network structures characterized by dyadic connections across the years. This fails to capture interactions among groups of vertices adequately \cite{sanatkar2015epidemic,pare2022multilayer}. Given the increasing complexity of social contagion processes, the presence of intricate group dynamics is undeniable and cannot be ignored. This motivates us to take into consideration higher-order networks, the most notable examples of which are \textit{simplicial complexes} and general \textit{hypergraphs}. Simplicial complexes are important special cases of hypergraphs, which contain all nonempty subsets of hyperedges as hyperedges \cite{battiston2020networks}. These high-order networks introduce hyperedges that encompass more than two agents to accurately describe the group interactions prevalent in various disciplines such as biology, communication, and sociology. Following this trend, some outstanding accomplishments on the SIS social contagion processes with higher-order networks and their bi-virus versions have been reported in \cite{iacopini2019simplicial,li2021contagion,chowdhary2021simplicial,gracy2023competitive}. However, in the existing body of research literature, few studies formally analyze the dynamic behavior of general multi-group compartmental epidemic models on general hypergraphs. In addition, the significant bifurcation conditions transitioning from the healthy-state domain to the bi-stable domain, and ultimately, to the endemic domain, are not rigorously provided except for in\cite{cisneros2021multigroup,gracy2023competitive,cui2023general}. Multigroup SIS epidemics with simplicial and
higher-order interactions are studied and the appearance conditions of the bistable domain are elucidated by employing the stability analysis of the mean-field theory in \cite{cisneros2021multigroup}. As pointed out in \cite{gracy2023competitive}, scholars have explored the spread of two competing viruses over a network of population nodes, accounting for pairwise and higher-order interactions within and between the population nodes. Nevertheless, to the best of our knowledge, the current research work on
SIS social contagion processes modeling using hypergraphs has been confined to the continuous-time domain, with limited attention
given to discrete-time SIS social contagion processes with general hypergraphs. Considering the nature of the sampled data, which is collected daily, it is extremely crucial to consider the stability issue for the discrete-time SIS social contagion processes with higher-order networks.

Inspired by the aforementioned analysis, the goal of this work is to study the issue of stability for networked SIS social contagion processes with hypergraphs. Our main contributions are two folds: (1) The discrete-time mean-field network SIS model is constructed based on its exact $2^n$-state Markov chain counterparts; Through Monte Carlo simulations, we show that these two models share similar performance on large-scale general directed hypergraphs with high connectivity.  (2) As a first effort, we provide a bifurcation analysis of the equilibrium of the discrete-time SIS model on general directed hypergraphs with second-order third-order hyperedges with the help of tensors. We further extend the results to SIS social contagion processes with multiple arbitrary high-order hyperedges and their bi-virus competing scenario over hypergraphs. (3) We propose many novel conclusions that are different from and stronger than those of the continuous-time counterpart \cite{cisneros2021multigroup,gracy2024networked,cui2023general}. Under novel and mild conditions and after knowing an equilibrium is locally stable, the domain of attraction can be directly calculated from both network topology and transition rates.

\textit{Notations:} The notations in the work are standard. $\mathsf{E}\{\cdot\}$ refers to the expectation. $\mathsf{diag}\{\cdot\}$ denotes a block diagonal matrix.  Given a square matrix $M\in \mathbb{R}^{n\times n}$, $\rho(M)$ is the spectral radius of $M$, which is the largest absolute value of the eigenvalues of $M$.  $s(M)$ symbols the largest real part among the eigenvalues of $M$. For vector $a\in \mathbb{R}^n$, $a_i$ is the $i$-th element of the vector $a$  for all $i = 1, \ldots , n$. For any two vectors $a, b \in \mathbb{R}^{n}$, $a \gg (\ll)b$ signifies that $a_i > (<)b_i$ for all $i = 1, \ldots , n$. $a > (<)b$ indicates that $a_i \geq (\leq)b_i$, for all $i = 1, \ldots , n$ and $a = b$. $a \geq (\leq)b$ represents that $a_i\geq (\leq)b_i$, for all $i = 1, \ldots, n$ or $a = b$. These component-wise comparisons are also applicable to matrices with the same dimension. The matrix $I$ means the identity matrix with the appropriate dimension. The vector \textbf{1 (0)} stands for the column vector or matrix of all ones (zeros) with suitable dimensions. 

\section{Preliminaries and Model Setup}

~~~~Firstly, weighted and directed hypergraph is introduced as a framework to model complex body interactions of any order, particularly in social contexts. Secondly, a Markov chain model is proposed to simulate the SIS social contagion process on this weighted and directed hypergraph, capturing the dynamics of how individuals transition between susceptible and infected states. Additionally, its mean-field model is constructed using mean-field theory to provide a macroscopic perspective on the contagion process. Finally, the discussion begins by examining SIS social contagion with both 2nd-order and 3rd-order hyperedges to establish foundational understanding, followed by an exploration of the rationale behind extending the bifurcation analysis to more general higher-order SIS social contagion, which aims to capture more intricate patterns of influence within social networks.

\subsection{Weighted directed hypergraph and tensor}
~~~~Consider a weighted and directed hypergraph $\mathcal{W}\!=\!\{\mathcal{N}, \mathcal{E},\mathcal{A}\}$, where $\mathcal{N}\!=\!\{1, 2, \ldots, n\}$ and $\mathcal{E}\!=\!\{E_1,E_2,\ldots,E_n\}$ 
$\subseteq \mathcal{N}\times\mathcal{N}$ signify the sets of vertices and hyperedges, respectively.
An ordered pair $E = (\mathcal{X}, \mathcal{Y})$ of disjoint subsets of vertices is adopted to portray the hyperedge of the weighted and directed hypergraph, where $\mathcal{X}$ is the tail of $E$ while $\mathcal{Y}$ is its head. Based on the physical interpretation of a spreading process scenario, our assumption is that each hyperedge is characterized by a single tail and either one or multiple heads. $\mathcal{A}=\{A_{2}, A_{3},\ldots, A_{n}\}$ represents the set of gathering the weights associated with all hyperedges, where $A_{\imath}$, $\imath=2,3,\ldots,n$, aggregates the weights of the $\imath$-order hyperedges, each of which connects $\imath$ vertices \cite{zhang2017second}. More specifically, for $i,j,k\in \mathcal{N},$ $A_3=[A_{ijk}]\geq 0$ collects the weights of all third-order hyperedges, and $A_{ijk}>0$ stands for the weight of the hyperedge where $i$ is the tail and $j, k$ are its heads.  If all hyperedges in the hypergraph are of the second order, then the hypergraph reduces to a standard graph. For the sake of brevity, within this paper, we employ the term "weight" (e.g., $A_{\bullet}$) to signify the corresponding hyperedge.

To enhance the representation and manipulation of multidimensional data, and to optimize computational efficiency, we introduce the concept of tensors from the perspectives of network model modeling and proof technique\cite{chen2024tensor,cui2024discrete1,da2018tensor}. The order of a tensor refers to the total number of its dimensions, with each dimension called a mode. A $k$th-order tensor is typically denoted by $A \in \mathbb{R}^{n_1\times n_2\times\ldots\times n_k}$. Hence, 
scalars $s \in \mathbb{R}$ are zero-order tensors, vectors $\textbf{v} \in \mathbb{R}^n$ are first-order tensors and matrices 
$M \in \mathbb{R}^{n\times m}$ are second-order tensors. 
A tensor is termed cubical if all its modes have equal size, denoted as $A \in \mathbb{R}^{n\times n\times\ldots\times n}$. We further express a $k$-th order $n$-dimensional cubical tensor
 as $A \in \mathbb{R}^{n\times n\times\ldots\times n}=\mathbb{R}^{[k,n]}$. Throughout the remainder of this paper, unless otherwise specified, a tensor will always refer to a cubical tensor. A cubical tensor is called supersymmetric if $A_{i_1i_2\ldots i_k}$ is invariant 
under any permutation of the indices. For example, a third-order tensor $A  \in \mathbb{R}^{n\times n\times n}$ is 
supersymmetric if it satisfies the following condition: 
$A_{i_1i_2i_3}=A_{i_1i_3i_2}=A_{i_2i_1i_3}=A_{i_2i_3i_1}=A_{i_3i_1i_2}=A_{i_3i_2i_1}$. The diagonal entries of a tensor are the entries with all the same index, for example, $A_{ii\ldots i}$. All other entries are called off-diagonal entries. A cubical tensor is called diagonal if $A_{i_1 i_2\ldots i_k} = 0$ except $i_1 = i_2 =\ldots= i_k$. The identity tensor $\mathcal{I}=(\gamma_{i_1i_2\ldots i_k})$ is defined as 
\begin{align*} 
\gamma_{i_1i_2\ldots i_k}=\left\{
  \begin{array}{ll}
1  \ \ \ \ \ \mathsf{if} \ i_1=i_2=\ldots=i_k,\\
0  \ \ \ \ \ \mathsf{otherwise}.
  \end{array}
\right.
\end{align*}

In our paper, inspired by Definition 1 in \cite{chen2022explicit}, we define a $k$th-order $n$-dimensional tensor $A\in\mathbb{R}^{[k,n]}$
 as almost symmetric if it is symmetric only with respect to its last $k-1$ modes. For example, a third-order tensor $A  \in \mathbb{R}^{n\times n\times n}$ is almost
symmetric if it satisfies the following condition: 
$A_{i_1i_2i_3}=A_{i_1i_3i_2}$. The definition in \cite{chen2022explicit} concerns symmetry with respect to the first $k-1$ modes. Nevertheless, we require symmetry concerning the last $k-1$ modes. These two definitions are consistent; the difference lies in using different tensor-vector multiplication notation. This notation we use, as shown below, is widely applied in the relevant literature on tensor algebra \cite{chauhan2024almost,chen2024tensor}. For a tensor $A \in \mathbb{R}^{[k,n]}$ and a vector $x\in \mathbb{R}^n $, the tensor-vector product $Ax$ is an order $k-1$ dimension $n$ tensor, whose component is given by $$(Ax)_{i_1i_2\ldots i_{k-1}}=\sum_{i_k=1}^{n}A_{i_1\ldots i_{k-1}i_k}x_{i_k}.$$ 
Similarly, the component of the tensor-vector product to the power $k-2$, $Ax^{k-2}$ is defined as $$(Ax^{k-2})_{i_1i_2}=\sum_{i_3,\ldots, i_k=1}^{n}A_{i_1i_2i_3\ldots i_k}x_{i_3}\ldots x_{i_k}.$$ The tensor-vector product to the power $k-1$, $Ax^{k-1}$ yields a vector, whose $i$-th component is given by $$(Ax^{k-1})_{i_1}=\sum_{i_2,\ldots,i_k=1}^{n}A_{i_1i_2\ldots i_k}x_{i_2}\ldots x_{i_k}.$$
 The component of the tensor-vector product to the power $k$ is defined as $$Ax^{k}=x^{\top}(Ax^{k-1})=\sum_{i_1,\ldots,i_k=1}^{n}A_{i_1i_2\ldots i_k}x_{i_1}\ldots x_{i_k}.$$ For a matrix $R\in\mathbb{R}^{n\times n},$ The component of the matrix-tensor product is as follows: $$(RA)_{i_1i_2\ldots i_k}=\sum_{j=1}^{n}R_{ij}A_{ji_2\ldots i_k}.$$ Recall that a tensor $A\in \mathbb{R}^{[k,n]}$ is called diagonally dominant if
$$|A_{ii\ldots i}| \geq \sum_{
(i2,\ldots,i_k)\neq(i,\ldots,i)}
|A_{ii_2\ldots i_k}|$$ for all $i = 1, 2, \ldots , n;$
and called strictly diagonally dominant if
$$|A_{ii\ldots i}| > \sum_{
(i2,\ldots,i_k)\neq(i,\ldots,i)}
|A_{ii_2\ldots i_k}|$$ for all $i = 1, 2, \ldots , n$ in \cite{chauhan2024almost}. A tensor $A = (A_{i_1\ldots i_k}) \in \mathbb{R}^{[k,n]}$ is called reducible if there is
a nonempty proper index subset $N_1 \subset \{1,\ldots, n\}$ such that
$A_{i_1\ldots i_k} = 0, \forall i_1 \in N_1, \forall i_2,\ldots,i_k \notin N_1.$
If $A$ is not reducible, then we call $A$ irreducible. A tensor with all non-negative entries is called a non-negative tensor.

\begin{lemma}\cite{chen2022explicit}\label{lemma_1}
Given a one dimensional homogenous polynomial function $f(x(t))$: $\mathbb{R}^n\rightarrow\mathbb{R}.$ It can be uniquely determined by $Ax^m$, $m$ is the order of $A$, $A$ is super-symmetric. Given a $n$-dimensional homogenous polynomial function $g(x(t))$: $\mathbb{R}^n\rightarrow\mathbb{R}^n.$ It can be uniquely determined by $Ax^{m-1}$, $m$ is the order of $A$, $A$ is almost symmetric.
\end{lemma}

\subsection{Network Model}

~~~~In the SIS social contagion processes, each agent of the weighted and directed hypergraph $\mathcal{W}$ is in one of the two compartments: susceptible (S) and infected (I), at each time instance. A susceptible agent may undergo infection when surrounded by infected agents. The agent's neighborhood is defined by the weighted and directed hypergraph $\mathcal{W}$, representing the contact network. Alongside the infection process, there is also a curing process. An infected agent transitions to susceptibility with a curing rate.

From a microscopic perspective, the network state transitions in the SIS manner can be modeled as a Markov process with $2^n$ states in continuous time, where $n$ represents the number of agents in hypergraph $\mathcal{W}$. Specifically, we denote the random variable $X_i(t)$ as the state of agent $i$ at time instance $t$. Moreover, $X_i(t)=0$ if agent $i$ is in state S and $X_i(t)=1$ if agent $i$ is in state I. As mentioned in \cite{cui2023general}, there are various propagation rules in the SIS social contagion processes, however, by reweighting these rules, they can be equivalently represented by the following propagation rule: for a $\imath$-order hyperedge ($\imath=2,3,\ldots,n$), infection of the tail agent may occur if and only if the remaining $\imath-1$ head agents have been infected. Similar to \cite{sahneh2013generalized}, it holds for the infection process that  for a sufficiently small time $\Delta t$,
\begin{align}
\mathsf{Prob}\{X_i(t+\Delta t)=1|X_i(t)=0\}
=&\sum_{j\in N_{2}^{i}}\mu_iA_{ij}X_j(t)\Delta t +\sum_{j,k\in N_{3}^{i}}\mu_{i3}A_{ijk}X_j(t)X_k(t)\Delta t +\ldots\notag\\ &+\sum_{j,k,l,\ldots\in N_{n}^{i}}\mu_{in}A_{ijkl\ldots}X_j(t)X_k(t)X_l(t)\ldots\Delta t+o(\Delta t)
\end{align}
where the infection rates $\mu_i,\mu_{i3},\ldots,\mu_{in} $ are the intrinsic parameter of agent $i$ to describe how likely $i$ is infected when the spreading rule is satisfied; whereas, the curing process is assumed to be passive with the rate $\delta_{i}$. The notation $N_{2}^{i}$ denotes the set including all the
conventional edges (pairwise interactions) in hypergraph $\mathcal{W}$, with the agent $i$ serving as the tail. Similarly, for $\imath=3,\ldots,n,$ the notation $N_{\imath}^{i}$ describes the set encompassing all the $\imath$-order hyperedges in hypergraph $\mathcal{W},$
where the agent $i$ is the tail. $o(\Delta t)$ is higher-order infinitesimal of $\Delta t$. Namely, $\lim_{\Delta t\rightarrow 0}\frac{o(\Delta t)}{\Delta t}=0$.

For the curing process, it holds that
\begin{align}
&\mathsf{Prob}\{X_i(t+\Delta t)=0|X_i(t)=1\}
=\delta_iX_i(t)\Delta t+o(\Delta t)
\end{align}

We now apply the mean-field approximation, leading to
\begin{align}
&\sum_{j\in N_{2}^{i}}\mu_iA_{ij}X_j(t)= \sum_{j\in N_{2}^{i}}\mu_iA_{ij}\mathsf{E}\{X_j(t) \}   \notag\\
&\sum_{j,k\in N_{3}^{i}}\mu_{i3}A_{ijk}X_j(t)X_k(t)= \sum_{j,k\in N_{3}^{i}}\mu_{i3}A_{ijk}\mathsf{E}\{X_j(t)X_k(t)\}    \notag\\
&\ \ \ \ \ \ \ \ \ \ \ \ \ \ \ \ \ \ \ \ \vdots  \notag\\
&\sum_{j,k,l,\ldots\in N_{n}^{i}}\mu_{in}A_{ijkl\ldots}X_j(t)X_k(t)X_l(t)\ldots=\sum_{j,k,l,\ldots\in N_{n}^{i}}\mu_{in}A_{ijkl\ldots}\mathsf{E}\{X_j(t)X_k(t)X_l(t)\ldots\}
\end{align}
This approximation is valid and accurate if the states
of the neighbors are sufficiently independent and the
number of the in-neighbors of agent $i$ is large. Then, when $\Delta t\rightarrow 0$, we get that
 \begin{align*}
 \frac{d\mathsf{E}\{X_i(t)\}}{dt}
 =&\mathsf{E}\{(\sum_{j\in N_{2}^{i}}\mu_iA_{ij}X_j(t)\!+\!\sum_{j,k\in N_{3}^{i}}\mu_{i3}A_{ijk}X_j(t) X_k(t) 
 \!+\!\ldots 
\!+\!\sum_{j,k,l,\ldots\in N_{n}^{i}}\mu_{in}A_{ijkl\ldots}X_j(t)\\
&\times X_k(t) X_l(t)\ldots) (1-X_i(t))-\delta_i X_i(t)\}.
\end{align*}

 By ignoring all covariance and letting $x_i(t)=\mathsf{Prob}\{X_i(t)=1\}=\mathsf{E}\{X_i(t)\}$, we get
 \begin{align} \label{continuous-time}
 \dot{x}_i(t)=&-\delta_ix_i(t)+(1-x_i(t))\sum_{j\in N_{2}^{i}}\mu_iA_{ij}x_j(t)
 +(1-x_i(t))\sum_{j,k\in N_{3}^{i}}\mu_{i3}A_{ijk}x_j(t)x_k(t)+\ldots \notag\\
 &+(1-x_i(t))\sum_{j,k,l\in N_{n}^{i}}\mu_{in}A_{ijkl\ldots}x_j(t)x_k(t)x_l(t)\ldots
 \end{align}
In the process of mean-field approximation, these random variables are assumed to be mutually independent, meaning that all covariance can be neglected. From this trend, according to this equation $\mathsf{E}\{X_{i}(t)X_{j}(t)\}=\mathsf{E}\{X_{i}(t)\}\mathsf{E}\{X_{j}(t)\}+\mathsf{cov}\{X_{i}(t), X_{j}(t)\}$, their joint expectation equals the product of their individual expectations. As is elaborated in \cite{PhysRevE.91.032812}, this kind of approximation is typically sufficiently accurate for large-scale networks, bearing in mind the Central Limit Theorem. The approximation error is usually tolerable in a network with several hundred agents.

 By applying Euler's method \cite{atkinson1991introduction}, we calculate the values $x_i(t+1)$ by making tangent line approximations:
\begin{align} \label{discrete-time}
 x_i(t+1)=&x_i(t)+h\left(-\delta_ix_i(t)+(1-x_i(t))\sum_{j\in N_{2}^{i}}\mu_iA_{ij}x_j(t)
 +(1-x_i(t))\sum_{j,k\in N_{3}^{i}}\mu_{i3}A_{ijk}x_j(t)x_k(t)+\ldots\right. \notag\\
 &\left.+(1-x_i(t))\sum_{j,k,l\in N_{n}^{i}}\mu_{in}A_{ijkl\ldots}x_j(t)x_k(t)x_l(t)\ldots\right),
 \end{align}
 where $h$ is the sampling parameter $(h>0)$.

\begin{remark}
Although continuous system (\ref{continuous-time}) is widely utilized, the investigation of discrete system (\ref{discrete-time}) is important for the advancement of sophisticated algorithms and the theoretical underpinnings in the realms of digital computation, data processing, control theory, and network analysis.
\end{remark}

\subsection{The network SIS social contagion processes involving both 2nd-order and 3rd-order hyperedges}
Let us now consider higher-order SIS social contagion processes on a hypergraph structure up to hyperedges of degree 3 (three agents involved in a hyperedge): for $i\in\mathcal{N},$
\begin{align} \label{3_agents-hyperedge_SIS_model}
 x_i(t+1)=&x_i(t)+h\left(-\delta_ix_i(t)+(1-x_i(t))\sum_{j\in N_{2}^{i}}\beta_{ij}x_j(t)
 +(1-x_i(t))\sum_{j,k\in N_{3}^{i}}\beta_{ijk}x_j(t)x_k(t)\right),
\end{align}
where for $i,j,k\in\mathcal{N},$ $\beta_{ij}=\mu_{i}A_{ij},\beta_{ijk}=\mu_{i3}A_{ijk}$.

We can further write it in tensor form as
\begin{align} \label{matrix_form}
 x(t+1)=x(t)+h\left(-\mathcal{D}x(t)+(I-\mathsf{diag}(x(t)))(\mathcal{B}x(t)+\mathcal{H}x^2(t))\right),
 \end{align}
 where $x(t)=[x_1(t) \ \ x_2(t) \ \ \ldots \ \ x_n(t)]^{\top},
\mathcal{D}=\mathsf{diag}([\delta_1 \ \ \delta_2 \ \ \ldots \ \  \delta_n]^{\top}),
\mathcal{B}=\left[\beta_{ij}\right]_{n\times n}\in\mathbb{R}^{[2,n]},$ and
$\mathcal{H}=\left[\beta_{ijk}\right]_{n\times n\times n}\in\mathbb{R}^{[3,n]},$ is almost symmetric tensor which can be uniquely chosen according to Lemma \ref{lemma_1}.

\begin{remark}
Generally speaking, $\mathcal{H}$ is not symmetric. However, from the system \eqref{3_agents-hyperedge_SIS_model}, we know that  $\beta_{ijk}x_j(t)x_k(t)+\beta_{ikj}x_k(t)x_j(t)\!=\!\frac{\beta_{ijk}+\beta_{ikj}}{2}x_j(t)x_k(t)+\frac{\beta_{ijk}+\beta_{ikj}}{2}x_k(t)x_j(t)$. In other words, this asymmetrical term $\mathcal{H}$ can be reweighted using simple mathematical calculation to make it almost symmetric.
\end{remark}

Before proceeding further, the following advantageous assumptions and lemma are borrowed to ensure that our model is well-defined and deduce the main results.

\begin{assumption} \label{assum_1}
For all $i\in\mathcal{N},$ $x_i(0)\in[0,1].$
\end{assumption}

\begin{assumption} \label{assum_2}
$\mathcal{D}$ is a positive diagonal matrix. $\mathcal{B}$ is an irreducible non-negative tensor. $\mathcal{H}$ is a non-negative tensor.
\end{assumption}

\begin{assumption} \label{assum_3}
(a) For all $i\in\mathcal{N},$ $h\delta_i\leq1$ and $h(\sum_{j\in\mathcal{N}}\beta_{ij}+\sum_{j,k\in\mathcal{N}}\beta_{ijk})\leq1$.
(b) For all $i\in\mathcal{N},$ $h(\delta_i+\sum_{j\in\mathcal{N}}\beta_{ij}+\sum_{j,k\in\mathcal{N}}\beta_{ijk})<1$.
\end{assumption}

\begin{lemma}\label{lemma_2}
For the system in (\ref{matrix_form}), under the conditions of Assumptions 1, 2, and 3a, $x_i(t)\in[0,1]$, for all $i\in\mathcal{N}$ and $t\geq0.$
\end{lemma}
{\bf proof} Firstly, for all $i\in\mathcal{N}$, the system (\ref{3_agents-hyperedge_SIS_model})  can be re-established as
\begin{align} \label{3_agents-hyperedge_SIS_model_1}
 x_i(t+1)=&x_i(t)\upsilon_1+(1-x_i(t))\upsilon_2,
\end{align}
where $\upsilon_1=1-h\delta_i$ and $\upsilon_2=h\left(\sum_{j\in N_{2}^{i}}\beta_{ij}x_j(t)+\sum_{j,k\in N_{3}^{i}}\beta_{ijk}x_j(t)x_k(t)\right).$

From (\ref{3_agents-hyperedge_SIS_model_1}), it is easy to see that $x_i(t+1)$ is a convex combination of $\upsilon_1$ and $\upsilon_2.$ Then, by mathematical induction, it holds that $x_i(t)\in[0,1]$, for all $i\in\mathcal{N}$ and $t\geq0.$ More specifically, for $t=0$ and recalling Assumption 1, $x_i(1)\in [0,1]$, for all $i\in\mathcal{N}$. In the sequel, assuming that at some time $t$, $x_i(t)\in[0,1]$, for all $i\in\mathcal{N}$. Combining Assumptions 2 and 3a, $\upsilon_1\in[0,1]$ and $\upsilon_2\in[0,1]$. Thus, one can gain that $x_i(t+1)\in[0,1]$, for all $i\in\mathcal{N}.$ {\hfill$\square$}


\begin{remark}
 The dynamics~\eqref{3_agents-hyperedge_SIS_model} is able to characterize not only epidemic spreading but also other dissemination processes, especially in social networks, e.g., the spreading of misinformation. In social contagion processes, group effects beyond pairwise connections play an important role. The proposed model involves hyperedges to model the higher-order interactions and is the basis for mathematical analysis and further extensions.

Based on the fact that the variable $x_i(t)$ represents the infection probability of agent $i$ or the infection proportion of the group $i$, it is natural to constrain the initial values of this SIS social contagion processes  (\ref{3_agents-hyperedge_SIS_model}) within the interval $[0,1]$. Considering the physical meaning and to ensure the well-posedness of the basic reproduction number defined later, the healing rates must be strictly positive, and the infection rates must be non-negative. With $h$ sufficiently small, suitable values of $\delta_i$, $\beta_{ij}$ and $\beta_{ijk}$ can be identified to satisfy Assumption 3a. Assumption 3b is stricter compared to Assumption 3a and contributes to the analysis of the existence and stability of the endemic equilibrium point in Theorem 2. Lemma 1 articulates a property, indicating that the set $[0,1]^n$ is positively invariant concerning the SIS social contagion processes  (\ref{3_agents-hyperedge_SIS_model}). This lemma also implies that the SIS social contagion processes are well-defined. Involving only pairwise interactions, the irreducible matrix $\mathcal{B}$ as stated in Assumption 2 implies that the underlying graph of this weighted and directed hypergraph is strongly connected, which is crucial for subsequent stability analysis of equilibrium points and has been widely applied \cite{pare2018analysis,cui2023general}. Given the reality of globalization, Assumption 2 is quite natural.
\end{remark}

\section{Analytical Results}
~~~~This section is devoted to analyzing the characteristics of discrete-time SIS models (\ref{matrix_form}) and their different behavior will be established: healthy-state behavior, bi-stable behavior, and endemic behavior. In greater detail, Proposition 1 and Theorem 1 delineate the novel sufficient conditions for the emergence of healthy-state behavior. Meanwhile, Proposition 2 and Theorem 2 reveal that, in contrast to the classical SIS model, the introduction of higher-order interaction terms leads to a novel dynamical phenomenon termed bi-stable behavior. This behavior entails the coexistence of both a healthy state  (zero equilibrium point)  and an endemic equilibrium  (non-zero equilibrium point), both of which are locally asymptotically stable. Proposition 3 and Theorem 3 provide novel sufficient conditions for the occurrence of endemic behavior. More importantly, considering the local stability of equilibrium, the regions of attraction for the health state and the endemic equilibrium are also provided in Theorem 1 and Corollary 1.

\begin{proposition} \label{th1}
Consider the system (\ref{matrix_form}). Given Assumptions 1, 2, and 3a hold, the healthy state always exists and is globally exponentially stable if
$\rho(\mathcal{D}^{-1}\mathcal{B}+\mathcal{D}^{-1}\mathcal{H}z)<1,$ with $z\in\mathbb{R}^n$, for $i\in\mathcal{N},$ $z_i=1$ if $ \mathcal{H}_{ijk}\neq 0$ for $j,k\in\mathcal{N}$ and $z_i=0$ otherwise.
\end{proposition}
{\bf proof} See Appendix 9.1.

Distinct from the SIS model on graphs with only dyadic connections, bi-stability could exist in the SIS model on hypergraphs. In this case, the disease-free and endemic equilibria both exist and are locally stable.  

\begin{proposition}\label{th2}(Bi-stability)
Consider the system (\ref{matrix_form}) under Assumptions 1, 2 and 3b. If $\rho(I-h\mathcal{D}+h\mathcal{B})<1$ and   $\theta\!=\!\mbox{min}_{i\in\mathcal{N},H_i\neq \textbf{0}_{n\times n}}$
$\{(2\mathcal{D}^{-1}\mathcal{B}z+\mathcal{D}^{-1}\mathcal{H}z^2)_i\}\geq4$, then the
healthy state is locally asymptotically stable and there exists an endemic equilibrium point $\bar{x}\gg\textbf{0}_n$ such that $\bar{x}_i\geq\frac{1}{2}$ for any $i\in\mathcal{N}$ such that $ \mathcal{H}_{ijk}\neq 0$ for $j,k\in\mathcal{N}$, which is locally asymptotically stable.
\end{proposition}
{\bf proof} See Appendix 9.2.

\begin{remark}
In this study, the significance of bistability cannot be overstated, and for the first time in discrete settings, we have provided the conditions under which it arises. There are similarities between Proposition 1 and Proposition 2 in that both entail $\rho(I-h\mathcal{D}+h\mathcal{B})$ being less than one, which is solely associated with ordinary graphs modeling pairwise relationships. More specifically, according to Assumptions 1 and 2, $\rho(\mathcal{D}^{-1}\mathcal{B}+\mathcal{D}^{-1}\mathcal{H}z)\geq\rho(\mathcal{D}^{-1}\mathcal{B})$, which implies that if $\rho(\mathcal{D}^{-1}\mathcal{B}+\mathcal{D}^{-1}\mathcal{H}z)$ $<1$, then $\rho(\mathcal{D}^{-1}\mathcal{B})<1$, or equivalently $\rho(I-h\mathcal{D}+h\mathcal{B})<1$. However, there are distinctions to note. Firstly, Proposition 2 requires Assumption 3b, which is beneficial for the stability analysis of the endemic equilibrium. Secondly, Proposition 2 introduces condition ($\theta\geq 4$) related to higher-order edges, thereby verifying the pivotal role that hypergraphs play in the emergence of bistability behavior. In other words, bi-stability is a contagion process that only exists in high-order networks, symbolizing the impact of higher-order structures on attraction domains corresponding to two classes of equilibrium points.
\end{remark}

So far, the focus has been on situations where $\rho(I-h\mathcal{D}+h\mathcal{B})<1$. Considering that $\rho(I-h\mathcal{D}+h\mathcal{B})>1$, an endemic behavior emerges, meaning that the endemic equilibria demonstrate global exponential stability.

\begin{proposition}\label{th3}
Consider the system (\ref{matrix_form}) under Assumptions 1, 2 and 3b. If $\rho(I-h\mathcal{D}+h\mathcal{B})>1$, and $\beta_{ijk}$ is sufficiently small for any $i,j,k\in\mathcal{N}$, then the
healthy state is unstable, and there exists a unique endemic equilibrium point $\bar{x}\gg\textbf{0}_n$, which is globally exponentially stable.
\end{proposition}
{\bf proof} See Appendix 9.3.

\begin{remark}
When $\rho(I-h\mathcal{D}+h\mathcal{B}) > 1$, it triggers a global endemic equilibrium, which is different from the previous case where it is less than 1. Thus, this value of $\rho(I-h\mathcal{D}+h\mathcal{B})$ is extremely important and can serve as an indicator of whether an epidemic will break out. In the SIS social contagion processes on conventional network structures \cite{cui2024discrete}, where only healthy-state behavior and endemic behavior are observed, scholars utilize the number $\rho(\mathcal{D}^{-1}\mathcal{B})$ as the basic reproduction number, i.e., the average number of individuals infected by a single infected person. As its value is associated with the existence and stability of both the healthy state and endemic equilibrium, it has been used to assess whether the SIS social contagion processes are in an endemic behavior. This number also plays a similar role in higher-order SIS social contagion processes. Notably, based on assumptions 1, and 2, $\rho(\mathcal{D}^{-1}\mathcal{B})\leq (>) 1$  is equivalent to $\rho(I-h\mathcal{D}+h\mathcal{B})\leq (>) 1$. This essentially allows us to view the number $\rho(I-h\mathcal{D}+h\mathcal{B})$ as the reproduction number, with the threshold set at 1.
\end{remark}

The properties of irreducible non-negative tensors, as presented in \cite{cui2024discrete1}, can be further utilized to study SIS social contagion processes over hypergraphs. Then, some novel sufficient conditions for the aforementioned behaviors will be proposed.

\begin{theorem}\label{tensor_th1}
Consider the system (\ref{matrix_form}) under Assumptions 1, 2, and 3a, and suppose that the tensor $\mathcal{H}$ is irreducible. If $\sum_{j\in\mathcal{N}}\beta_{ij}<\delta_i<\sum_{j\in\mathcal{N}}\beta_{ij}+\sum_{j,k\in\mathcal{N}}\beta_{ijk}$ for all $i\in\mathcal{N},$ the healthy state is asymptotically stable within a domain of attraction $\max_{i\in\mathcal{N}}|x_i(0)|<\alpha_1=\min_{i\in\mathcal{N}}\frac{\delta_i-\sum_{j\in\mathcal{N}}\beta_{ij}}{\sum_{j,k\in\mathcal{N}}\beta_{ijk}}$. If $\delta_i>\sum_{j\in\mathcal{N}}\beta_{ij}+\sum_{j,k\in\mathcal{N}}\beta_{ijk}$ for all $i\in\mathcal{N},$ the healthy state is globally asymptotically stable. 
\end{theorem}
{\bf proof}

Notice that 
\begin{align} 
 x(t+1)\leq(I-h\mathcal{D}+h\mathcal{B})x(t)+h\mathcal{H}x^2(t).
\end{align}
Recalling that $\delta_i>\sum_{j\in\mathcal{N}}\beta_{ij}$ for all $i\in\mathcal{N}$ and Assumption 2, it is easy to achieve that $\max_{i\in\mathcal{N}}\sum_{j\in\mathcal{N}}|I-h\mathcal{D}+h\mathcal{B}|_{ij}<1.$ In virtue of Assumption 1, 2, 3a and the irreducible tensor $\mathcal{H},$ $I-h\mathcal{D}+h\mathcal{B}$ and $ h\mathcal{H}$ are irreducible. Thus,
following Theorem 3 and Corollary 1 in \cite{cui2024discrete1} and comparison principle, one can obtain that the healthy state is asymptotically stable within a domain of attraction $\max_{i\in\mathcal{N}}|x_i(0)|<\alpha_1$. Should $\delta_i<\sum_{j\in\mathcal{N}}\beta_{ij}+\sum_{j,k\in\mathcal{N}}\beta_{ijk}$ for all $i\in\mathcal{N}$ , then $\alpha_1<1$; conversely, should  $\delta_i>\sum_{j\in\mathcal{N}}\beta_{ij}+\sum_{j,k\in\mathcal{N}}\beta_{ijk}$ for all $i\in\mathcal{N}$ , then $\alpha_1>1$. Thus, one completes the proof. 
{\hfill$\square$}
\\
\begin{remark}\label{remark_5}
Proposition 1 presents a sufficient condition for the global stability of the healthy state. However, this condition is excessively intricate and lacks clear physical meaning. Conversely, the condition in Theorem 1 is more intuitively meaningful in a physical sense. Specifically, the global stability of the healthy state is assured if the recovery rate $(\delta_i)$ of each node surpasses the aggregate of all corresponding second-order and third-order infection rates $(\beta_{ij}$ and  $\beta_{ijk})$. Under a novel condition, determining a conservative domain of attraction from the system’s parameters is also significant after establishing the local stability of equilibrium. Based on the Perron-Frobenius theorem for irreducible non-negative matrix and the Gershgorin circle theorem, if $\delta_i>\sum_{j\in\mathcal{N}}\beta_{ij}$, we have $\rho(I-h\mathcal{D}+h\mathcal{B})<1-h\delta_i+h\sum_{j\in\mathcal{N}}\beta_{ij}<1$, which shows that the health state is locally asymptotically stable based on Proposition 2. Additionally, if $\delta_i>\sum_{j\in\mathcal{N}}\beta_{ij}+\sum_{j,k\in\mathcal{N}}\beta_{ijk}$, it holds that $\rho(\mathcal{D}^{-1}\mathcal{B}+\mathcal{D}^{-1}\mathcal{H}z)<\delta_i^{-1}\sum_{j\in\mathcal{N}}\beta_{ij}+\delta_i^{-1}\sum_{j,k\in\mathcal{N}}\beta_{ijk}<1$, which indicates the connectivity of Theorem 1 and Proposition 1. From Theorem 1, it can be seen that the global stability condition of the healthy state is more stringent than the local stability condition. Moreover, note that when compared to that of SIS models on conventional graphs \cite{pare2018analysis}, the global stability condition of the healthy state is observed to vary depending on the network structure. This underscores the impact that different network configurations exert.
\end{remark}

Subsequently, leveraging tensor notation, we shall delineate the endemic behavior of the system (\ref{matrix_form}). Initially, by defining the error variable $y(t)=x(t)-\bar{x}$ where $\bar{x}$ is the endemic equilibrium of the system (\ref{matrix_form}), the error dynamics of the system (\ref{matrix_form}) can be reformulated as
\begin{align} \label{error_dynamics1}
y(t+1)=&(I-h\mathcal{D})y(t)+h(I-\mathsf{diag}(\bar{x}))(\mathcal{B}y(t)+\mathcal{H}y^2(t)+2\mathcal{H}\bar{x}y(t))\notag\\
&-h\mathsf{diag}(y(t))(\mathcal{B}y(t)+\mathcal{B}\bar{x}+\mathcal{H}y^2(t)+2\mathcal{H}\bar{x}y(t)+\mathcal{H}\bar{x}^2).
\end{align}
Then, since the Taylor series expansion of any polynomial at a given point is the polynomial itself, the Taylor expansion for the system (\ref{matrix_form}) with respect to an endemic equilibrium $\bar{x}$ can be rewritten as
\begin{align} \label{error_dynamics2}
y(t+1)=&(I-h\mathcal{D}+h(I-\mathsf{diag}(\bar{x}))(\mathcal{B}+2\mathcal{H}\bar{x})\notag\\
&\ \ \ \ -h\mathsf{diag}(\mathcal{B}\bar{x}+\mathcal{H}\bar{x}^2))y(t)+o(y(t)).
\end{align}
where $o(y(t))$ is higher-order infinitesimal of $y(t)$.
\begin{theorem}\label{tensor_th2}
Consider the system (\ref{error_dynamics2}) under Assumptions 1, 2, and 3a. Then, the following statements hold: 
\begin{itemize}  
\item[i)] (Bistability) If there exists an endemic equilibrium $\bar{x}\geq\frac{2}{3}\textbf{1}_n$ such that $\delta_i>(1-\bar{x}_i)\sum_{j\in\mathcal{N}}(\beta_{ij}+2\sum_{k\in\mathcal{N}}\beta_{ijk}\bar{x}_k)$ and $\bar{x}_i<\frac{2\sum_{j,k\in\mathcal{N}}\beta_{ijk}\bar{x}_k}{2\sum_{j,k\in\mathcal{N}}\beta_{ijk}\bar{x}_k+\sum_{j\in\mathcal{N}}\beta_{ij}}$ for all $i\in\mathcal{N}$, then the health state and the endemic equilibrium $\bar{x}$ for the system (\ref{matrix_form}) are both locally asymptotically stable. 
\item[ii)] If there exists an endemic equilibrium $\bar{x}\geq\frac{2}{3}\textbf{1}_n$ such that $\delta_i>(1-\bar{x}_i)\sum_{j\in\mathcal{N}}(\beta_{ij}+2\sum_{k\in\mathcal{N}}\beta_{ijk}\bar{x}_k)$ and \\$\frac{2\sum_{j,k\in\mathcal{N}}\beta_{ijk}\bar{x}_k}{2\sum_{j,k\in\mathcal{N}}\beta_{ijk}\bar{x}_k+\sum_{j\in\mathcal{N}}\beta_{ij}}<\bar{x}_i<1$ for all $i\in\mathcal{N}$, then the endemic equilibrium $\bar{x}$ for the system (\ref{matrix_form}) is locally asymptotically stable. 
\end{itemize}
\end{theorem}
{\bf Proof}
Since $I-h\mathcal{D}+h(I-\mathsf{diag}(\bar{x}))(\mathcal{B}+2\mathcal{H}\bar{x})$ is irreducible nonnegative matrix regarding the Assumption \ref{assum_2}, \ref{assum_3}a, and Lemma \ref{lemma_2}, let $\lambda:=\rho(I-h\mathcal{D}+h(I-\mathsf{diag}(\bar{x}))(\mathcal{B}+2\mathcal{H}\bar{x}))$ be its eigenvalue based on the Perron-Frobenius theorem (Theorem 8.4.4) in \cite{horn2012matrix}. It follows from the Gershgorin circle theorem (Theorem 6.1.1) in \cite{Horn_Johnson_2012} that
$|\lambda-(1-h\delta_i+h(1-\bar{x}_i)(\beta_{ii}+2\sum_{k\in\mathcal{N}}\beta_{iik}\bar{x}_k))|\leq (1-\bar{x}_i)\sum_{j\in\mathcal{N},j\neq i}(h\beta_{ij}+2h\sum_{k\in\mathcal{N}}\beta_{ijk}\bar{x}_k))$ for all $i\in\mathcal{N}$. Combining the fact that there exists an endemic equilibrium $\bar{x}\geq\frac{2}{3}\textbf{1}_n$ such that $\delta_i>(1-\bar{x}_i)\sum_{j\in\mathcal{N}}(\beta_{ij}+2\sum_{k\in\mathcal{N}}\beta_{ijk}\bar{x}_k)$, one attains that $\lambda<1-h\delta_i+(1-\bar{x}_i)\sum_{j\in\mathcal{N}}(h\beta_{ij}+2h\sum_{k\in\mathcal{N}}\beta_{ijk}\bar{x}_k))<1$. Note that the Jacobian of the system (\ref{error_dynamics2}) at the health state is given by $I-h\mathcal{D}+h(I-\mathsf{diag}(\bar{x}))(\mathcal{B}+2\mathcal{H}\bar{x})-h\mathsf{diag}(\mathcal{B}\bar{x}+\mathcal{H}\bar{x}^2)$. Besides, since $\textbf{0}_{n\times n}\leq I-h\mathcal{D}+h(I-\mathsf{diag}(\bar{x}))(\mathcal{B}+2\mathcal{H}\bar{x})-h\mathsf{diag}(\mathcal{B}\bar{x}+\mathcal{H}\bar{x}^2)\leq I-h\mathcal{D}+h(I-\mathsf{diag}(\bar{x}))(\mathcal{B}+2\mathcal{H}\bar{x})$, it is generated from Lemma 2 in \cite{cui2024metzler} that $\rho (I-h\mathcal{D}+h(I-\mathsf{diag}(\bar{x}))(\mathcal{B}+2\mathcal{H}\bar{x})-h\mathsf{diag}(\mathcal{B}\bar{x}+\mathcal{H}\bar{x}^2))\leq\lambda=\rho(I-h\mathcal{D}+h(I-\mathsf{diag}(\bar{x}))(\mathcal{B}+2\mathcal{H}\bar{x}))<1$. Thus, the health state of the system (\ref{error_dynamics2}) is locally asymptotically stable according to Theorem 42 in Section 5.9 of the work \cite{vidyasagar2002nonlinear}. In other words, the endemic equilibrium $\bar{x}$ of the system (\ref{matrix_form}) is locally asymptotically stable. 

 On the other hand, considering the condition $\bar{x}_i<$ $\frac{2\sum_{j,k\in\mathcal{N}}\beta_{ijk}\bar{x}_k}{2\sum_{j,k\in\mathcal{N}}\beta_{ijk}\bar{x}_k+\sum_{j\in\mathcal{N}}\beta_{ij}}$ for all $i\in\mathcal{N}$, one can get that $\delta_i>\sum_{j\in\mathcal{N}}\beta_{ij}$, which yields that the health state of the system (\ref{matrix_form}) is locally asymptotically stable from Theorem \ref{tensor_th1}. However, if $\frac{2\sum_{j,k\in\mathcal{N}}\beta_{ijk}\bar{x}_k}{2\sum_{j,k\in\mathcal{N}}\beta_{ijk}\bar{x}_k+\sum_{j\in\mathcal{N}}\beta_{ij}}<\bar{x}_i<1$, then the above conclusion will not hold. This ends the proof.  {\hfill$\square$}

\begin{remark}
Under the premise of establishing the error dynamics \eqref{error_dynamics2} using tensor algebra, we adopt a completely new proof method to consider the endemic equilibrium for the system \eqref{matrix_form}. Through this new method, we obtain novel conditions related to each agent. Compared to the spectral radius conditions in Propositions 1-3, our proposed conditions are more applicable to large-scale network systems. The conditions of Theorem \ref{tensor_th2} (ii) allow the inequality $\rho(I-h\mathcal{D}+h\mathcal{B})>1$ to appear. In other words, considering $\rho(I-h\mathcal{D}+h\mathcal{B})>1$, the endemic equilibrium $\bar{x}$ for the system (\ref{matrix_form}) will be stable only if its value is sufficiently large and the health state for the system (\ref{matrix_form}) is unstable based on Proposition \ref{th3}. The conditions of Theorem \ref{tensor_th2} (i) indicate that $\rho(I-h\mathcal{D}+h\mathcal{B})<1$, recalling Remark \ref{remark_5}.
\end{remark}

Theorem \ref{tensor_th2} provides the local stability conditions of the endemic equilibrium $\bar{x}$ for the system (\ref{matrix_form}) but does not specify the domain of attraction. The following theorem addresses this issue by not only providing the domain of attraction for the endemic equilibrium $\bar{x}$ but also considering the stability of the endemic equilibrium $\bar{x}$ from a different perspective of irreducible nonnegative tensors, as discussed in detail in \cite{cui2024discrete1}.

\begin{theorem}\label{tensor_th21}
Consider the system (\ref{error_dynamics1}) under Assumptions 1, 2, and 3b, and suppose that the tensor $\mathcal{H}$ is irreducible. If there exists an endemic equilibrium $\bar{x}$ such that $\delta_i>(1-\bar{x}_i)\sum_{j\in\mathcal{N}}(\beta_{ij}+2\sum_{k\in\mathcal{N}}\beta_{ijk}\bar{x}_k)-(\sum_{j\in\mathcal{N}}\beta_{ij}\bar{x}_j+\sum_{j,k\in\mathcal{N}}\beta_{ijk}\bar{x}_j\bar{x}_k)$ for all $i\in\mathcal{N}$, then the endemic equilibrium $\bar{x}$ for the system (\ref{matrix_form}) is asymptotically stable with a domain of attraction $\max_{i\in\mathcal{N}}|x_i(0)|<\alpha_2=\min_{i\in\mathcal{N}}\frac{-\overline{\mathcal{K}}_2+\sqrt{\overline{\mathcal{K}}_2^2-4\overline{\mathcal{K}}_3(\overline{\mathcal{K}}_1-1)}}{2\overline{\mathcal{K}}_3}$ where $\overline{\mathcal{K}}_1=\sum_{j\in\mathcal{N}}|\mathcal{K}_1|_{ij}, \!\overline{\mathcal{K}}_2\!=\!\sum_{j,k\in\mathcal{N}}|\mathcal{K}_2|_{ijk}, \!\overline{\mathcal{K}}_3\!=\!\sum_{j,k,\ell\in\mathcal{N}}|\mathcal{K}_1|_{ijk\ell}$, $\mathcal{K}_1=I-h\mathcal{D}+h(I-\mathsf{diag}(\bar{x}))(\mathcal{B}+2\mathcal{H}\bar{x})-h\mathsf{diag}(\mathcal{B}\bar{x}+\mathcal{H}\bar{x}^2), \mathcal{K}_2=h(I-\mathsf{diag}(\bar{x}))\mathcal{H}-h(\widetilde{\mathcal{B}}+2\widetilde{\mathcal{H}}\bar{x}),$  $\mathcal{K}_3=-h\widetilde{\mathcal{H}}$, $\widetilde{\mathcal{B}}=\left[\widetilde{\beta}_{ijk}\right]_{n\times n\times n}\in\mathbb{R}^{3,n}$, and $\widetilde{\mathcal{H}}=\left[\widetilde{\beta}_{ijk\ell}\right]_{n\times n\times n\times n}\in\mathbb{R}^{4,n}$  with $\widetilde{\beta}_{iik}=\beta_{ik}$, $\widetilde{\beta}_{iik\ell}=\beta_{ik\ell}$ for all $i, k, \ell\in \mathcal{N}$ and all other entries are zero.    \end{theorem}
{\bf Proof} 
In light of the computational analysis, the error dynamics (\ref{error_dynamics1}) can be reformulated as follows: $y(t+1)=\mathcal{K}_1y(t)+\mathcal{K}_2y^2(t)+\mathcal{K}_3y^3(t)$. Recalling Assumption \ref{assum_2} that the matrix $\mathcal{B}$ is irreducible and since the tensor $\mathcal{H}$ is irreducible, all the tensors $\mathcal{K}_1, \mathcal{K}_2, \mathcal{K}_3$ are irreducible. Note that $\mathcal{K}_1$ is nonnegative according to Assumptions \ref{assum_2}, \ref{assum_3}b, and Lemma \ref{lemma_2}. Then, from the inequality $\delta_i>(1-\bar{x}_i)\sum_{j\in\mathcal{N}}(\beta_{ij}+2\sum_{k\in\mathcal{N}}\beta_{ijk}\bar{x}_k)-(\sum_{j\in\mathcal{N}}\beta_{ij}\bar{x}_j+\sum_{j,k\in\mathcal{N}}\beta_{ijk}\bar{x}_j\bar{x}_k)$, one can get $\max_{i}(\sum_{j\in\mathcal{N}}|\mathcal{K}_1|_{ij})=\max_{i}(\sum_{j\in\mathcal{N}}(\mathcal{K}_1)_{ij})=\max_{i}(1-h\delta_i-h(\sum_{j\in\mathcal{N}}\beta_{ij}\bar{x}_j+\sum_{j,k\in\mathcal{N}}\beta_{ijk}\bar{x}_j\bar{x}_k)+h(1-\bar{x}_i)\sum_{j\in\mathcal{N}}(\beta_{ij}+2\sum_{k\in\mathcal{N}}\beta_{ijk}\bar{x}_k) )<1$. Therefore, according to Corollary 2 in \cite{cui2024discrete1}, one can obtain that the healthy state of the error dynamics (\ref{error_dynamics1}) is asymptotically stable within a domain of attraction $\max_{i\in\mathcal{N}}|x_i(0)|<\alpha_2$. This yields that the endemic equilibrium $\bar{x}$ for the system (\ref{matrix_form}) is asymptotically stable with a domain of attraction $\max_{i\in\mathcal{N}}|x_i(0)|<\alpha_2$.    {\hfill$\square$}
\begin{remark}\label{remark_8}
Under the assumption that the endemic equilibrium exists, Theorem \ref{tensor_th21} constructs the domain of attraction for the local stability of the endemic equilibrium. If $\overline{\mathcal{K}}_1+\overline{\mathcal{K}}_2+\overline{\mathcal{K}}_3<1$, then $\alpha_2>1$. In other words, according to Theorem \ref{tensor_th21}, the endemic equilibrium $\bar{x}$ for the system (\ref{matrix_form}) is globally asymptotically stable. Moreover, compared to the stability conditions of the endemic equilibrium in Theorem \ref{tensor_th2}, the stability conditions provided in Theorem \ref{tensor_th21} are more precise and less conservative. This is because the proof of Theorem \ref{tensor_th2} does not utilize the scaling condition $\rho(I-h\mathcal{D}+h(I-\mathsf{diag}(\bar{x}))(\mathcal{B}+2\mathcal{H}\bar{x})-h\mathsf{diag}(\mathcal{B}\bar{x}+\mathcal{H}\bar{x}^2))<\rho(I-h\mathcal{D}+h(I-\mathsf{diag}(\bar{x}))(\mathcal{B}+2\mathcal{H}\bar{x}))$. Theorem \ref{tensor_th2} and Theorem \ref{tensor_th21}, assuming that an endemic equilibrium exists,
guarantee its local stability. Following this trend, the issue of the existence of the endemic equilibrium still needs to be addressed. Then, if $\mathcal{B}$ is a strictly diagonally dominant matrix and $\delta_i<\beta_{ii}-\sum_{j\in\mathcal{N},j\neq i}\beta_{ij}$ for all $i\in\mathcal{N}$, one can realize that $\rho(I-h\mathcal{D}+h\mathcal{B})>1-h\delta_i+h\beta_{ii}-h\sum_{j\in\mathcal{N},j\neq i}\beta_{ij}>1$, regarding the Gershgorin circle theorem. Subsequently, there exists an endemic equilibrium point $\bar{x}\gg \textbf{0}_n$ based on the Proposition \ref{th3}.
\end{remark}

Combining the aforementioned theorems, these conclusions constitute a bifurcation analysis process.

\section{Parameter Learning}
In this section, we provide a learning technique to obtain the heterogeneous spreading parameters for the system (\ref{matrix_form}). In detail, the parameter identification problem is formally presented as follows

\begin{problem}\label{Parameter_Identification_th}
Given the system (\ref{matrix_form}) under Assumptions 1-3 with $n > 1$. Assume that $h$ and$A_{ij}, A_{ijk}, x_i(t), i,j,k\in\mathcal{N}, t \geq 0$ are known. Calculate the spreading parameters $\delta_i,\mu_i,\mu_{i3}$ of agent $i$.
\end{problem}

Rewrite the system (\ref{matrix_form}) at $t=q,q+1,\ldots,q+m$ with $q,m \geq 0$ as follows 
\begin{align}\label{Parameter_Identification}
x_i(t+1)-x_i(t)=-hx_i(t)\delta_i+((1-x_i(t))\sum_{j\in\mathcal{N}}A_{ij}x_j(t))\mu_i +((1-x_i(t))\sum_{j,k\in\mathcal{N}}A_{ijk}x_j(t)x_k(t))\mu_{i3}.
\end{align}
We can construct the following linear equations 
\begin{align} \label{eq:para_learning}
  \begin{bmatrix}
    x_i(q+1)-x_i(q) \\
    x_i(q+2)-x_i(q+1) \\
    \vdots \\
    x_i(q+m)-x_i(q+m-1) \\
  \end{bmatrix}=\Phi_i\begin{bmatrix}
                            \delta_i \\
                            \mu_i \\
                            \mu_{i3}. \\
                          \end{bmatrix},
\end{align}
where
\begin{align*}
\Phi_i\!=\!\left[\begin{smallmatrix}
  -hx_i(q) & h(1-x_i(q))\sum_{j\in\mathcal{N}}A_{ij}x_j(q) & h(1-x_i(q))\sum_{j,k\in\mathcal{N}}A_{ijk}x_j(q)x_k(q) \\
  -hx_i(q+1) & h(1-x_i(q+1))\sum_{j\in\mathcal{N}}A_{ij}x_j(q+1) & h(1-x_i(q+1))\sum_{j,k\in\mathcal{N}}A_{ijk}x_j(q+1)x_k(q+1) \\
  \vdots & \vdots & \vdots \\
  -hx_i(q+m\!-\!1) & h(1-x_i(q+m\!-\!1))\sum_{j\in\mathcal{N}}A_{ij}x_j(q+m\!-\!1) & h(1-x_i(q+m\!-\!1))\sum_{j,k\in\mathcal{N}}A_{ijk}x_j(q+m\!-\!1)x_k(q+m\!-\!1) \\
\end{smallmatrix}\right],
\end{align*}
which can be further streamlined into the following form $\eta_i=\Phi_i\theta_i^*$. Thus, we transform the parameter learning problem of the system (\ref{matrix_form}) into solving the linear equations~\eqref{eq:para_learning}. Clearly, it can be solved by directly using the least-square technique. More specifically, the process of identifying parameters can be structured as a constrained optimization problem in the following manner:
\begin{equation*}\theta_i^*=\mathop{\arg}\mathop{\min}_{\theta_i} \frac{1}{2}||\Phi_i\theta_i-\eta_i||_2^2 .
\end{equation*}
If $\Phi_i$ is full column rank, the solution to the spread parameters for each agent $i$ can be learned uniquely. This can be simply satisfied by choosing a long period $m$ in the transient process.

\section{Extension to competitive bi-virus SIS social contagion processes with high-order interactions}
In most real-world propagation scenarios, it is not uncommon to encounter the simultaneous presence of multiple viruses. For instance, in the context of product adoption in markets and the dissemination of opinions on social networks, various social contagion processes often coexist, potentially engaging in either cooperative \cite{gracy2022modeling} or competitive interactions \cite{darabi2014competitive}. This section delves into the simplest case of bi-virus competition as a typical example. Specifically, suppose that the infection by one virus excludes the possibility of co-infection by another virus \cite{cui2024discrete2}. Under this assumption, the system (\ref{3_agents-hyperedge_SIS_model}) can be straightforwardly extended into a bi-virus competing SIS social contagion processes as 
\begin{align} \label{bi_virus_1}
\left\{
  \begin{array}{ll}
x_{1i}(t+1)=&x_{1i}(t)+h\left(-\delta_{1i}x_{1i}(t)+(1-x_{1i}(t)-x_{2i}(t))\left(\sum_{j\in N_{2}^{i}}\beta_{1,ij}x_{1i}(t)\right.\right.\\
 &\left. \left.\ \ \ \ \ \ \ +\sum_{j,k\in N_{3}^{i}}\beta_{1,ijk}x_{1j}(t)x_{1k}(t)\right)\right),\\
x_{2i}(t+1)=&x_{2i}(t)+h\left(-\delta_{2i}x_{2i}(t)+(1-x_{1i}(t)-x_{2i}(t))\left(\sum_{j\in N_{2}^{i}}\beta_{2,ij}x_{2i}(t)\right.\right.\\
 &\left. \left.\ \ \ \ \ \ \  +\sum_{j,k\in N_{3}^{i}}\beta_{2,ijk}x_{2j}(t)x_{2k}(t)\right)\right).
  \end{array}
\right.
\end{align}
where $i,j,k\in\mathcal{N},$ and $\ell=1,2,$ $\beta_{\ell,ij}=\mu_{\ell i}A_{\ell,ij},\beta_{\ell,ijk}=\mu_{\ell i3}A_{\ell,ijk}$. These symbols retain the same physical meaning in the bi-virus model as in the single-virus model (\ref{3_agents-hyperedge_SIS_model}).

For $\ell=1,2$, the system (\ref{bi_virus_1}) can be written in tensor form as
\begin{align} \label{bi_virus_2}
x_{[\ell]}(t+1)=x_{[\ell]}(t)+h\left(-\mathcal{D}_{\ell}x_{[\ell]}(t)+(1-\mathsf{diag}(x_{[1]}(t))-\mathsf{diag}(x_{[2]}(t)))\left(\mathcal{B}_{\ell}x_{[\ell]}(t)
 +\mathcal{H}_{\ell}x_{[\ell]}^2(t)\right)\right).
\end{align}
 where $x_{[\ell]}(t)=[x_{\ell 1}(t) \ \ x_{\ell 2}(t) \ \ \ldots \ \ x_{\ell n}(t)]^{\top},
\mathcal{D}_{\ell}=\mathsf{diag}([\delta_{\ell 1} \ \ \delta_{\ell 2} \ \ \ldots \ \  \delta_{\ell n}]^{\top}),
\mathcal{B}_{\ell}=\left[\beta_{\ell,ij}\right]_{n\times n}\in\mathbb{R}^{[2,n]},
\mathcal{H}_{\ell}=\left[\beta_{\ell,ijk}\right]_{n\times n\times n}\in\mathbb{R}^{[3,n]}.
$  

In order to ensure that the bi-virus competing system (\ref{bi_virus_2}) are well-defined, the bi-virus version of the assumptions~\ref{assum_1}-\ref{assum_3} are presented. Based on these assumptions, some basic properties of the bi-virus dynamics are obtained.
\begin{assumption} \label{bi_virus_assumption1}
 For all $\ell=1,2,i\in\mathcal{N},$  $ x_{\ell i}(0),1-x_{1i}(0)- x_{2i}(0)\in[0,1].$ 
\end{assumption}

\begin{assumption}\label{bi_virus_assumption2}
For all $\ell=1,2,$ $\mathcal{D}_{\ell}$ is a positive diagonal matrix. $\mathcal{B}_{\ell}$ is an irreducible non-negative tensor. $\mathcal{H}_{\ell}$ is a non-negative tensor.
\end{assumption}

\begin{assumption}\label{bi_virus_assumption3}
(a) For all $\ell=1,2,i\in\mathcal{N},$ $h\delta_{\ell i}\leq1$ and $\sum_{\ell=1}^2h(\sum_{j\in\mathcal{N}}\beta_{\ell,ij}+\sum_{j,k\in\mathcal{N}}\beta_{\ell,ijk})\leq1$.
(b) For all $i\in\mathcal{N},$ $h(\delta_{\ell i}+\sum_{\ell=1}^2(\sum_{j\in\mathcal{N}}\beta_{\ell,ij}+\sum_{j,k\in\mathcal{N}}\beta_{\ell,ijk}))<1$.
\end{assumption}

\begin{lemma} \label{bi_virus_lemma1}
For the system (\ref{bi_virus_2}), under the conditions of Assumptions \ref{bi_virus_assumption1}, \ref{bi_virus_assumption2}, and \ref{bi_virus_assumption3}a, $x_{\ell i}(t),1-x_{1i}(t)- x_{2i}(t)\in[0,1]$, for all $i\in\mathcal{N},\ell=1,2,$ and $t\geq0.$
\end{lemma}
{\bf Proof} 
For $t=0$ and by Assumption \ref{bi_virus_assumption1}, we find $\sum_{\ell=1}^{2}x_{\ell i}(1)\in[0,1]$. In what follows, suppose that $\sum_{\ell=1}^{2}x_{\ell i}(t).$ Then, based on the system (\ref{bi_virus_1}), we have
\begin{align*}
\sum_{\ell=1}^{2}x_{\ell i}(t+1)&=\sum_{\ell=1}^{2}x_{\ell i}(t)(1-h\delta_{\ell i})+(1-\sum_{\ell=1}^{2}x_{\ell i}(t))h
 \sum_{\ell=1}^{2}(\sum_{j\in\mathcal{N}}\beta_{\ell,ij}x_{\ell j}(t)+\sum_{j,k\in\mathcal{N}}\beta_{\ell,ijk}x_{\ell j}(t)x_{\ell k}(t))\\
&\leq\sum_{\ell=1}^{2}x_{\ell i}(t)(1-h\min_{\ell=1,2}\{\delta_{\ell i}\})+(1-\sum_{\ell=1}^{2}x_{\ell i}(t)) h\sum_{\ell=1}^{2}(\sum_{j\in\mathcal{N}}\beta_{\ell,ij}x_{\ell j}(t)+\sum_{j,k\in\mathcal{N}}\beta_{\ell,ijk}x_{\ell j}(t)x_{\ell k}(t))\\
&= \Theta(t)
\end{align*}
By Assumption \ref{bi_virus_assumption3}a, we have $1-h\min_{\ell=1,2}\{\delta_{\ell i}\}\in[0,1)$. Then, according to Assumption \ref{bi_virus_assumption3}a and since $x_{\ell i}(t)\in[0,1]$,  $h\sum_{\ell=1}^{2}(\sum_{j\in\mathcal{N}}\beta_{\ell,ij}x_{\ell j}(t)+\sum_{j,k\in\mathcal{N}}\beta_{\ell,ijk}x_{\ell j}(t)x_{\ell k}(t))\in[0,1]$. Therefore, one has $\Theta(t)\in[0,1]$, which yields $\sum_{\ell=1}^{2}x_{\ell i}(t+1)\in[0,1]$. Then, by induction, we prove
that $\sum_{\ell=1}^{2}x_{\ell i}(t)\in[0,1]$.This yields $x_{\ell i}(t),1-x_{1i}(t)- x_{2i}(t)\in[0,1]$, for all $i\in\mathcal{N},\ell=1,2,$ and $t\geq0.$ {\hfill$\square$}

Lemma \ref{bi_virus_lemma1} implies that the set $\mathcal{S}=\{(x_{[1]},x_{[2]})|x_{[\ell]}\geq\textbf{0}_n,\ell=1,2,\sum_{\ell=1}^{2}x_{[\ell]}\leq\textbf{1}_n\}$ is positively invariant.

\begin{lemma} \label{bi_virus_lemma2}
For the system (\ref{bi_virus_2}), under the conditions of Assumptions \ref{bi_virus_assumption1}, \ref{bi_virus_assumption2}, and \ref{bi_virus_assumption3}a, if $(\bar{x}_{[1]},\bar{x}_{[2]})\in\mathcal{S}$ is an equilibrium of the system (\ref{bi_virus_2}), then for each $\ell=1,2$, either $\bar{x}_{[\ell]}=\textbf{0}_n$ or $\textbf{0}_n\ll\bar{x}_{[\ell]}\ll\textbf{1}_n$. Moreover, $\sum_{\ell=1}^{2}\bar{x}_{[\ell]}\ll\textbf{1}_n$.
\end{lemma}
{\bf Proof} Firstly, we prove that any non-zero
equilibrium $\bar{x}=(\bar{x}_{[1]},\bar{x}_{[2]})>0$ for the bivirus competing system (\ref{bi_virus_2}), satisfies $\textbf{0}_n\ll\bar{x}_{[\ell]}\ll\textbf{1}_n$ and $\sum_{\ell=1}^{2}\bar{x}_{[\ell]}\ll\textbf{1}_n$. Assuming that, for some $i\in\mathcal{N}$, we have $\sum_{\ell=1}^{2}\bar{x}_{\ell i}=1$. Then, based on the facts that $\delta_i>0$ in Assumption  \ref{bi_virus_assumption2} and $h>0$, $\sum_{\ell=1}^{2}\bar{x}_{\ell i}=1-h\delta_i<1$, which  is a contradiction. Then, we have $\sum_{\ell=1}^{2}\bar{x}_{[\ell]}\ll\textbf{1}_n$. Since $\bar{x}_{[1]}>0$, suppose that, for some $i\in\mathcal{N}$, we have $\bar{x}_{1 i}=0$. Then, one can get $\bar{x}_{1i}=(1-\bar{x}_{1i}-\bar{x}_{2i})(\sum_{j\in\mathcal{N}}\beta_{1,ij}\bar{x}_{1j}+\sum_{j,k\in\mathcal{N}}\beta_{1,ijk}\bar{x}_{1j}\bar{x}_{1k})>0$, which  is a contradiction. Then, we have $\bar{x}_{[1]}\gg0$. Similarly, $\bar{x}_{[2]}\gg0$. It is clear that $(\textbf{0}_n, \textbf{0}_n)$ is an equilibrium of the bivirus competing system (\ref{bi_virus_2}). Thus completing the proof.  {\hfill$\square$}

In what follows, based on the bi-virus competing system (\ref{bi_virus_2}), we present analytical results. Firstly, 
for bi-virus competing system on a conventional graph, if $\rho(I-h\mathcal{D}_{\ell}+h\mathcal{B}_{\ell})<1$, then the healthy
state is the unique equilibrium of system (\ref{bi_virus_2}), which is
globally asymptotically stable in Theorem 10 of the work \cite{cui2024discrete2}. However, when considering the bi-virus competition system on a hypergraph, this assertion does not necessarily hold. The following theorem reveals the conditions under which all viruses will eventually die out, leading to healthy-state behavior.

\begin{proposition}
Consider the system (\ref{bi_virus_2}). Given Assumptions \ref{bi_virus_assumption1}, \ref{bi_virus_assumption2}, and \ref{bi_virus_assumption3}a hold, the
healthy state $(\textbf{0}_n,\textbf{0}_n)$ always exists and is globally exponentially stable if
$\rho(\mathcal{D}_{\ell}^{-1}\mathcal{B}_{\ell}+\mathcal{D}_{\ell}^{-1}\mathcal{H}_{\ell}z_{[\ell]})<1$ for $\ell=1,2,$ with $z_{[\ell]}\in\mathbb{R}^n$, for $i\in\mathcal{N}$, $z_{\ell i}=1$ if $(\mathcal{H}_{\ell})_{ijk}\neq\textbf{0}$ for $j,k\in\mathcal{N}$ and $z_{\ell i}=0$ otherwise.
\end{proposition}
{\bf proof} See Appendix 9.4.

\begin{proposition}\label{tensor_th11}
Consider the system (\ref{bi_virus_2}). Given Assumptions \ref{bi_virus_assumption1}, \ref{bi_virus_assumption2}, and \ref{bi_virus_assumption3}a hold, and suppose that the tensor $\mathcal{H}_{\ell}$ is irreducible for each $\ell=1,2$. If $\sum_{j\in\mathcal{N}}\beta_{\ell,ij}<\delta_{\ell i}<\sum_{j\in\mathcal{N}}\beta_{\ell,ij}+\sum_{j,k\in\mathcal{N}}\beta_{\ell,ijk}$ for all $i\in\mathcal{N}, \ell=1,2$ the healthy state $(\textbf{0}_n,\textbf{0}_n)$ is asymptotically stable within a domain of attraction $\{\left(x_{[1]}(0),x_{[2]}(0)\right)|x_{[1]}(0)\in \mathcal{S}_1,x_{[2]}(0)\in \mathcal{S}_2\}$, where $\mathcal{S}_{\ell}=\Big\{x_{[\ell]}(0)|\max_{i\in\mathcal{N}}|x_{\ell i}(0)|$ $<\min_{i\in\mathcal{N}}\frac{\delta_{\ell i}-\sum_{j\in\mathcal{N}}\beta_{\ell,ij}}{\sum_{j,k\in\mathcal{N}}\beta_{\ell,ijk}}\Big\}$ is the domain of attraction for single virus $\ell$ for each $\ell=1,2$. If $\delta_{\ell i}>\sum_{j\in\mathcal{N}}\beta_{\ell,ij}+\sum_{j,k\in\mathcal{N}}\beta_{\ell,ijk}$ for all $i\in\mathcal{N},$ the healthy state $(\textbf{0}_n,\textbf{0}_n)$ is globally asymptotically stable.
\end{proposition}
{\bf Proof} See Appendix 9.5.

Sometimes, one virus may dominate the spreading process, leading to a "winner takes all" scenario. This means that after the competition, only one virus will survive while all others will die out. We refer to this equilibrium state as the dominant endemic equilibrium. The following theorem identifies a parameter regime that allows the three equilibria of the bi-virus competing system (\ref{bi_virus_2}), namely the healthy state and the two dominant endemic equilibria, to be simultaneously locally asymptotically stable.

\begin{proposition}\label{bi_virus_th2}(Multi-stability)
Consider the system (\ref{bi_virus_2}) under Assumptions \ref{bi_virus_assumption1}, \ref{bi_virus_assumption2}, and \ref{bi_virus_assumption3}. If for each $\ell=1,2,$ $\rho(I-h\mathcal{D}_{\ell}+h\mathcal{B}_{\ell})<1$ and   $\theta_{\ell}\!=\!\mbox{min}_{i\in\mathcal{N},H_{\ell i}\neq\textbf{0}_{n\!\times \!n}}$
$\{(2\mathcal{D}_{\ell}^{-1}\mathcal{B}_{\ell}z_{[\ell]}+\mathcal{D}_{\ell}^{-1}\mathcal{H}_{\ell}z_{[\ell]}^2)_i\}\geq4$, then the
healthy state $(\textbf{0}_n,\textbf{0}_n)$ is locally asymptotically stable and there exists single-virus endemic equilibrium point $\bar{x}_{[\ell]}\gg\textbf{0}_n$ corresponding to virus $\ell$ such that $\bar{x}_{\ell i}\geq\frac{1}{2}$ for any $i\in\mathcal{N}$ such that $(\mathcal{H}_{\ell})_{ijk}\neq\textbf{0}$ for $j,k\in\mathcal{N}$, which is locally asymptotically stable. Moreover,   the dominant endemic equilibria $(\bar{x}_{[1]},\textbf{0}_n)$ and $(\textbf{0}_n,\bar{x}_{[2]})$ are locally asymptotically stable.
\end{proposition}
{\bf proof} See Appendix 9.6.

\begin{remark}
Proposition \ref{bi_virus_th2} elucidates an interesting phenomenon by providing a parameter condition that allows three equilibria, namely the healthy state and two dominant endemic equilibria, to simultaneously exist and be stable. This phenomenon is unique to hypergraphs and does not occur in conventional graphs. It also extends the previously studied single-virus case, which permits the coexistence and stability of both the healthy state and the endemic equilibrium.  
\end{remark}

Following the trend, we obtain a different set of parameter conditions from those previously mentioned, ensuring the existence and stability of the dominant endemic equilibria, as well as the existence of coexisting equilibrium.

\begin{proposition}\label{bi_virus_th3}
Consider the system (\ref{bi_virus_2}) under Assumptions \ref{bi_virus_assumption1}, \ref{bi_virus_assumption2},  \ref{bi_virus_assumption3}b, and $\rho(I-h\mathcal{D}_{\ell}+h\mathcal{B}_{\ell})>1$ for $\ell=1,2.$ Then, the
healthy state $(\textbf{0}_n,\textbf{0}_n)$ is unstable, and there exists a single-virus endemic equilibrium point $\bar{x}_{[\ell]}\gg\textbf{0}_n$ corresponding to virus $\ell$. Moreover, the following statements hold:
i) if $\rho(I+h(-\mathcal{D}_{\ell}+(I-\textit{diag}\{\bar{x}_{[\imath]}\})\mathcal{B}_{\ell}))>1$ for $\ell,\imath=1,2,\ell\neq\imath,$  the dominant endemic equilibria $(\bar{x}_{[1]},\textbf{0}_n)$ and $(\textbf{0}_n,\bar{x}_{[2]})$ are unstable and there exists at least one coexisting equilibrium $(\hat{x}_{[1]},\hat{x}_{[2]})$ with $\textbf{0}_n\ll\hat{x}_{[1]},\hat{x}_{[2]}\ll\textbf{1}_n$ and $\hat{x}_{[1]}+\hat{x}_{[2]}\ll\textbf{1}_n$.
ii) if for $\ell=1,2,$ $\rho(\mathcal{D}_{\ell}^{-1}\mathcal{B}_{\ell}+\mathcal{D}_{\ell}^{-1}\mathcal{H}_{\ell}z_{[\ell]})<1$ and $\beta_{\ell,ijk}$ is sufficiently small for any $i,j,k\in\mathcal{N}$, then the dominant endemic equilibria $(\bar{x}_{[1]},\textbf{0}_n)$ and $(\textbf{0}_n,\bar{x}_{[2]})$ are globally exponentially stable with domain of attraction $\mathcal{S}\verb|\| \mathcal{L}_1$ and $\mathcal{S}\verb|\| \mathcal{L}_2,$ where $\mathcal{L}_1=\{(x_{[1]},x_{[2]}), x_{[1]}=\textbf{0}_n, x_{[2]}=\mathcal{S}\}$ and $\mathcal{L}_2=\{(x_{[1]},x_{[2]}), x_{[1]}=\mathcal{S}, x_{[2]}=\textbf{0}_n\}$.
\end{proposition}
{\bf proof} See Appendix 9.7.

It is noteworthy that in Proposition \ref{bi_virus_th3}, the proof of the existence for the coexistence equilibrium $(\hat{x}_{[1]},\hat{x}_{[2]})$ relies on the assumption that the dominant endemic equilibria are unstable. Building on this, we propose a different condition that ensures the existence of coexistence equilibrium $(\hat{x}_{[1]},\hat{x}_{[2]})$ while also allowing both the dominant endemic equilibria to be stable.

\begin{theorem} \label{bi_virus_th}
Consider the system (\ref{bi_virus_2}) under Assumptions \ref{bi_virus_assumption1}, \ref{bi_virus_assumption2},  \ref{bi_virus_assumption3}b. Suppose that there are dominant endemic equilibria $(\bar{x}_{[1]},\textbf{0}_n)$ and $(\textbf{0}_n,\bar{x}_{[2]})$ which are locally asymptotically stable. If $\rho(I+h(-\mathcal{D}_{\ell}+(I-\textit{diag}\{\bar{x}_{[\nu]}\})\mathcal{B}_{\ell}))>1$ for $\ell,\nu=1,2,\ell\neq\nu,$  there exists coexisting equilibrium $(\hat{x}_{[1]},\hat{x}_{[2]})$ with $\textbf{0}_n\ll\hat{x}_{[1]},\hat{x}_{[2]}\ll\textbf{1}_n$ and $\hat{x}_{[1]}+\hat{x}_{[2]}\ll\textbf{1}_n$.
\end{theorem}
{\bf Proof}
 If Assumptions \ref{bi_virus_assumption1}, \ref{bi_virus_assumption2}, and \ref{bi_virus_assumption3}b hold, then Assumptions 1 and 2
in \cite{gracy2024networked} hold. Since $I+h(-\mathcal{D}_{\ell}+(I-\textit{diag}\{\bar{x}_{[\nu]}\})\mathcal{B}_{\ell})$ is irreducible nonnegative
matrix, we have $s(I+h(-\mathcal{D}_{\ell}+(I-\textit{diag}\{\bar{x}_{[\nu]}\})\mathcal{B}_{\ell}))=\rho(I+h(-\mathcal{D}_{\ell}+(I-\textit{diag}\{\bar{x}_{[\nu]}\})\mathcal{B}_{\ell}))>1$ for $\ell,\nu=1,2,\ell\neq\nu,$ recalling the Perron Frobenius Theorem. This indicates that the conditions (i) and (ii) in Theorem 5.2 of \cite{gracy2024networked} hold. Then, by
Lemma 16 in \cite{cui2024discrete2}, the system \eqref{bi_virus_2} and its continuous-time model share
the same equilibria. Therefore, by Theorem 5.2 of \cite{gracy2024networked} and the fact that there are dominant endemic equilibria $(\bar{x}_{[1]},\textbf{0}_n)$ and $(\textbf{0}_n,\bar{x}_{[2]})$ which are locally asymptotically stable, there exists coexisting equilibrium $(\hat{x}_{[1]},\hat{x}_{[2]})$ with $\textbf{0}_n\ll\hat{x}_{[1]},\hat{x}_{[2]}\ll\textbf{1}_n$ and $\hat{x}_{[1]}+\hat{x}_{[2]}\ll\textbf{1}_n$. Then, one can complete the proof.  {\hfill$\square$}

\begin{remark}
Based on Assumptions \ref{bi_virus_assumption2} and \ref{bi_virus_assumption3}a, we have $I+h(-\mathcal{D}_{\ell}+(I-\textit{diag}\{\bar{x}_{[\imath]}\})\mathcal{B}_{\ell})$ is non-negative irreducible matrix. Then, recalling the  Gershgorin circle theorem, it is obvious that $\rho(I+h(-\mathcal{D}_{\ell}+(I-\textit{diag}\{\bar{x}_{[\imath]}\})\mathcal{B}_{\ell}))>1-h\delta_{\ell i}+h(1-\bar{x}_{\imath i})\beta_{\ell,ii}-h(1-\bar{x}_{\imath i})\sum_{j\in\mathcal{N}}\beta_{\ell,ij}.$ Therefore, if $\mathcal{B}_{\ell}$ is a strictly diagonally
dominant matrix and there exist dominant endemic equilibria $(\bar{x}_{[1]},\textbf{0}_n)$ and $(\textbf{0}_n,\bar{x}_{[2]})$ such that $\delta_{\ell i}<(1-\bar{x}_{\imath i})(\beta_{\ell,ii}-\sum_{j\in\mathcal{N},j\neq i}\beta_{\ell,ij})$ for all $i\in\mathcal{N}$, $\ell, \imath=1,2,\ell\neq\imath,$ we can get that $\rho(I+h(-\mathcal{D}_{\ell}+(I-\textit{diag}\{\bar{x}_{[\imath]}\})\mathcal{B}_{\ell}))>1$.
\end{remark}

\section{Extension to general higher-order SIS social contagion processes}
In the section, the SIS social contagion processes with multiple arbitrary high-order interactions are studied. It can be aware that our analysis in sections 3 and 4 is limited to interactions involving up to three bodies. Nevertheless, under suitable changes on the sufficient conditions of section 3, it is straightforward that the parallel outcomes to Proposition 1-3 will be obtained for SIS social contagion processes that extend beyond interactions among three bodies.
 
In conjugation with the system (\ref{discrete-time}), the  SIS social contagion processes can be rewritten as for all $i\in\mathcal{N}$ 
 \begin{align} \label{discrete-time1}
 x_i(t+1)=x_i(t)+h&\left(-\delta_ix_i(t)+(1-x_i(t))\sum_{j\in N_{2}^{i}}\mu_iA_{ij}x_j(t)
 +(1-x_i(t))\sum_{k=3}^{n}\mu_{ik}\right.\notag\\
 &\left.\times\sum_{i_1,i_2,\ldots, i_{k-1}\in N_{k}^{i}}A_{ii_1i_2\ldots i_{k-1}}x_{i_1}(t)x_{i_2}(t)\ldots x_{i_{k-1}}(t)\right),
 \end{align}
whose tensor form is as follows:
\begin{align} \label{matrix_form3}
 x(t+1)=x(t)+h\left(-\mathcal{D}x(t)+(I-\mathsf{diag}(x(t))) (\mathcal{B}x(t)+\sum_{k=3}^{n}\mathcal{F}_kx^{k-1}(t))\right),
 \end{align}
 where $\mathcal{F}_k=\left[\beta_{ii_1i_2\ldots i_{k-1}}\right]_{n\times n\times \ldots \times n}\in\mathbb{R}^{[k,n]}$ is almost symmetric tensor with $\beta_{ii_1i_2\ldots i_{k-1}}=\mu_{ik}A_{ii_1i_2\ldots i_{k-1}},$ for $k=3,4,\ldots,n.$

In order to ensure our model (\ref{matrix_form3}) is well-defined, a set of assumptions and lemma analogous to Section 2.3 should be introduced. Since the proof of Lemma 1 is essentially independent of any higher-order interactions, these assumptions and lemma can be extended to general higher-order SIS models.

\begin{assumption}\label{General_assumption1}
$\mathcal{F}_k$ is a non-negative tensor for $k=3,4,\ldots,n.$    
\end{assumption}

\begin{assumption}\label{General_assumption2}
(a) For all $i\in\mathcal{N},$ $h\delta_i\leq1$ and $h\left(\sum_{j\in\mathcal{N}}\beta_{ij}+\sum_{k=3}^{n}\sum_{i_1,i_2,\ldots, i_{k-1}\in N_{k}^{i}}\beta_{ii_1i_2\ldots i_{k-1}}\right)\leq1$.
(b) For all $i\in\mathcal{N},$ $h\left(\delta_i+\sum_{j\in\mathcal{N}}\beta_{ij}+\sum_{k=3}^{n}\sum_{i_1,i_2,\ldots, i_{k-1}\in N_{k}^{i}}\beta_{ii_1i_2\ldots i_{k-1}}\right)<1$.
\end{assumption}

\begin{lemma}\label{g_lemma1}
For the system in (\ref{matrix_form3}), under the conditions of Assumptions 1, 2, \ref{General_assumption1}, and \ref{General_assumption2}a, $x_i(t)\in[0,1]$, for all $i\in\mathcal{N}$ and $t\geq0.$
\end{lemma}

In what follows, the analytical results for the general higher-order SIS social contagion processes are presented.

\begin{proposition} \label{General_th1}
Consider the system (\ref{matrix_form3}). Given Assumptions 1, 2, \ref{General_assumption1}, and \ref{General_assumption2}a, the healthy state always exists and is globally exponentially stable if
$\rho(\mathcal{D}^{-1}\mathcal{B}+\mathcal{D}^{-1}\sum_{k=3}^n\mathcal{F}_k\tilde{z}^{k-2})<1,$ with $\tilde{z}\in\mathbb{R}^n$ where for $i\in\mathcal{N},$
$\tilde{z}_i=1$, if  $\sum_{i_1,i_2,\ldots, i_{k-1}\in N_{k}^{i}}(\mathcal{F}_k)_{ii_1i_2\ldots i_{k-1}}\neq 0$ for $k=3,4,\ldots,n$, and $\tilde{z}_i=0$ otherwise.
\end{proposition}

The proof for Proposition \ref{General_th1} is similar to Proposition 1 in section 3.

\begin{proposition}\label{General_th3}(Bi-stability)
Consider the system (\ref{matrix_form3}), under Assumptions 1, 2, \ref{General_assumption1}, and \ref{General_assumption2}b. If $\rho(I-h\mathcal{D}+h\mathcal{B})<1$ and   $\tilde{\theta}=\mbox{min}_{i\in\mathcal{N}}$
$\{(\mathcal{D}^{-1}(\mathcal{B}\tilde{z} +\sum_{k=3}^{n}(\frac{n-2}{n-1})^{k-2} \mathcal{F}_k\tilde{z}^{k-1})_i\}\geq n-1$, then the healthy state is locally asymptotically stable and there exists an endemic equilibrium point $\bar{x}\gg\textbf{0}_n$ such that $\bar{x}\geq\frac{n-2}{n-1}\textbf{1}_n$, which is locally asymptotically stable.
\end{proposition}
{\bf proof} See Appendix 9.9.

\begin{proposition}\label{General_th2}
Consider the system (\ref{matrix_form3}) under Assumptions 1, 2, \ref{General_assumption1}, and \ref{General_assumption2}b. If $\rho(I-h\mathcal{D}+h\mathcal{B})>1$, and for $k=3,4,\ldots,n,i\in\mathcal{N},i_1,i_2,$
$\ldots, i_{k-1}\in N_{k}^{i}$, $\beta_{ii_1i_2\ldots i_{k-1}}$ is sufficiently small, then the
healthy state is unstable and there exists an endemic equilibrium point $\bar{x}\gg\textbf{0}_n$, which is globally exponentially stable.
\end{proposition}

The proof for Proposition \ref{General_th2} is similar to Proposition 3 in section 3.

\begin{remark}
 In Propositions \ref{General_th1} and \ref{General_th2}, given $k$ equals 3, signifying that the SIS social contagion processes are solely focusing on the second and third-order hyperedges, such Propositions can be reduced to Propositions 1 and 3. Transitioning to Proposition \ref{General_th3}, it is ascertained that the value of $\tilde{\theta}$ fulfilling bistability is intrinsically linked not only to the propagation parameters and network structure but also to the network scale denoted as $n$. The value of the endemic equilibrium point is required to be at the minimum, equal to or greater than $\frac{n-2}{n-1}$. It is critical to highlight that the value of $\frac{n-2}{n-1}$ incrementally escalates with an increase in $n$, indicating an elevated difficulty in obtaining bistability with larger network scales. Should the system account for the presence of third-order hyperedges, the network scale should not be less than 3. With the $k$ value in Proposition \ref{General_th3} being 3, indicating the order of the hyperedge, and the $n$ value also equating to 3, representing the network scale, the conditions of Proposition \ref{General_th3} would align with those of Proposition 2.
\end{remark}

Next, we will explore the health state and the endemic equilibrium for the system (\ref{matrix_form3}) from various perspectives, and provide the corresponding domains of attraction.

\begin{theorem}
Consider the system (\ref{matrix_form3}) under Assumptions 1, 2, \ref{General_assumption1}, and \ref{General_assumption2}a. If the tensor $\mathcal{F}_k$ is irreducible for $k=3,4,\ldots,n,$ and $\delta_i>\sum_{j\in\mathcal{N}}\beta_{ij}$ for all $i\in\mathcal{N},$ the healthy state is asymtotically stable within a domain of attraction $\max_{i\in\mathcal{N}}|x_i(0)|<\min_{i\in\mathcal{N}}p_{i+}$, where $p_{i+}$ is the unique positive scalar solution of $\sum_{k=1}^{n-1}\sum_{\mathcal{I}_{k}}|(\mathcal{C}_{k})_{i,\mathcal{I}_{k}}|y^{k-1}=1$ corresponding to the index $i$ and $\mathcal{C}_1=I-h\mathcal{D}+h\mathcal{B}$,  $\mathcal{C}_{\jmath}=h\mathcal{F}_{\jmath+1},\jmath=2,3,\ldots,n-1$, $\mathcal{I}_1=\{i_{n-1}\},\mathcal{I}_2=\{i_{n-2},i_{n-1}\},\ldots,\mathcal{I}_{n-1}=\{i_{1},\ldots,i_{n-1}\}.$
\end{theorem}
{\bf proof}

Note that the system (\ref{matrix_form3}) can be written as $$x(t+1) \leq\left(I-h\mathcal{D}+h\mathcal{B}\right)x(t)+\sum_{k=3}^{n}h\mathcal{F}_kx^{k-1}(t)=\sum_{k=1}^{n-1}\mathcal{C}_kx^k(t).$$ Then, considering that $\delta_i>\sum_{j\in\mathcal{N}}\beta_{ij}$ for all $i\in\mathcal{N}$ and Assumption 2, we have $\max_{i\in\mathcal{N}}\sum_{j\in\mathcal{N}}|\mathcal{C}_1|_{ij}<1.$ In virtue of Assumptions 1, 2, \ref{General_assumption1}, and \ref{General_assumption2}a and the irreducible tensor $\mathcal{F}_k,$ $\mathcal{C}_1,\mathcal{C}_2,\ldots,\mathcal{C}_{n-1}$ are irreducible. Thus,
following Theorem 3 in \cite{cui2024discrete1} and comparison principle, one can show that the healthy state is asymptotically stable within a domain of attraction $\max_{i\in\mathcal{N}}|x_i(0)|<\min_{i\in\mathcal{N}}p_{i+}$. Thus, one completes the proof. 
{\hfill$\square$}

In what follows, the error dynamics for the system \eqref{matrix_form3} can be reorganized as
\begin{align}\label{error_general}
    y(t+1)=\mathcal{G}_1 y(t)+\mathcal{G}_2 y^2(t)+\ldots +\mathcal{G}_n y^n(t),
\end{align}
where 
\begin{align*}
\mathcal{G}_1&= I-h\mathcal{D}+h(I-\mathsf{diag}(\bar{x}))(\mathcal{B}+\sum_{k=3}^{n}\mathcal{F}_kC_{k-1}^{1}\bar{x}^{k-2})-h\mathsf{diag}(\mathcal{B}\bar{x}+\sum_{k=3}^{n}\mathcal{F}_k\bar{x}^{k-1}), \\
\mathcal{G}_2&=h(I-\mathsf{diag}(\bar{x}))\sum_{k=3}^{n}\mathcal{F}_kC_{k-1}^{2}\bar{x}^{k-3}-h\sum_{k=3}^{n}\widetilde{\mathcal{F}}_kC_{k-1}^{1}\bar{x}^{k-2},\\
\mathcal{G}_{i}&=\sum_{k=i+1}^{n}\left(h(I-\mathsf{diag}(\bar{x}))\mathcal{F}_kC_{k-1}^{i}\bar{x}^{k-1-i}-h\widetilde{\mathcal{F}}_kC_{k-1}^{i-1}\bar{x}^{k-i}\right)-h\widetilde{\mathcal{F}}_{i}  \ \ \mbox{for} \ \ i=3,4,\ldots,n-1,\\
\mathcal{G}_n&=-h\widetilde{\mathcal{F}}_n, \widetilde{\mathcal{F}}_k=\left[\widetilde{\beta}_{ii_1i_2\ldots i_{k}}\right]_{n\times n\times \ldots \times n}\in\mathbb{R}^{[k+1,n]} \ \ \mbox{for} \ \ k=1,2,\ldots,n, \widetilde{\beta}_{iii_2\ldots i_{k}}=\beta_{ii_2\ldots i_{k}},\\
\widetilde{\beta}&_{ii_1i_2\ldots i_{k}}=0 \ \ \mbox{for} \ \ i_1\neq i, i_1,i_2,\dots,i_k\in\mathcal{N},
\end{align*}
$C_{n}^{m}$ represents the combination number, indicating the number of ways to choose $m$ elements from a set of $n$ elements without considering the order. Please refer to Appendix 9.8 for the derivation process of the error dynamics \eqref{error_general}.

\begin{theorem}
Consider the system (\ref{error_general}) under Assumptions 1, 2, \ref{General_assumption1}, and \ref{General_assumption2}b, and suppose that the tensor $\mathcal{F}_k$ is irreducible for $k=3,4,\ldots,n,$ if there exists an endemic equilibrium $\bar{x}$ such that $\delta_i>\alpha_{3i}=(1-\bar{x}_i)(\sum_{j\in\mathcal{N}}\beta_{ij}+\sum_{k=3}^{n}\sum_{i_1,\ldots,i_{k-1}\in\mathcal{N}}C_{k-1}^{1}\beta_{ii_1\ldots i_{k-1}}\bar{x}_{i_2}\ldots\bar{x}_{i_{k-1}})-(\sum_{j\in\mathcal{N}}\beta_{ij}\bar{x}_j+\sum_{k=3}^{n}\sum_{i_1,\ldots,i_{k-1}\in\mathcal{N}}\beta_{ii_1\ldots i_{k-1}}\bar{x}_{i_1}\ldots\bar{x}_{i_{k-1}})$ for all $i\in\mathcal{N},$ the endemic equilibrium $\bar{x}$ for the system (\ref{matrix_form3}) is asymtotically stable within a domain of attraction $\max_{i\in\mathcal{N}}|x_i(0)|<\min_{i\in\mathcal{N}}p_{i+}$, where $p_{i+}$ is the unique positive scalar solution of $\sum_{k=1}^{n}\sum_{\mathcal{I}_{k}}|(\mathcal{G}_{k})_{i,\mathcal{I}_{k}}|y^{k-1}=1$ corresponding to the index $i$ and $\mathcal{I}_1=\{i_{n}\},\mathcal{I}_2=\{i_{n-1},i_{n}\},\ldots,\mathcal{I}_{n}=\{i_{1},\ldots,i_{n}\}.$
\end{theorem}
{\bf proof}

According to the inequality $\delta_i>\alpha_{3i}$ for all $i\in\mathcal{N}$, it is a straightforward calculation yields that $\max_{i}(\sum_{j\in\mathcal{N}}|\mathcal{G}_1|_{ij})<1$. In addition, recalling Assumption 2, \ref{General_assumption1}, and \ref{General_assumption2}b, and the irreducible tensor $\mathcal{F}_k$, one can obtain that the tensors $\mathcal{G}_1,\mathcal{G}_2, \ldots,\mathcal{G}_n $ are irreducible. Then, it follows from Theorem 3 in \cite{cui2024discrete1} that, the healthy state of the error dynamics (\ref{error_general}) is asymptotically stable within a domain of attraction $\max_{i\in\mathcal{N}}|x_i(0)|<\min_{i\in\mathcal{N}}p_{i+}$. Therefore, the endemic equilibrium $\bar{x}$ for the system (\ref{matrix_form3}) is asymptotically stable with a domain of attraction $\max_{i\in\mathcal{N}}|x_i(0)|<\min_{i\in\mathcal{N}}p_{i+}$.     {\hfill$\square$}

\begin{remark}
Note that if one of the following conditions is satisfied: $\delta_i>(1-\bar{x}_i)(\sum_{j\in\mathcal{N}}\beta_{ij}+\sum_{k=3}^{n}\sum_{i_1,\ldots,i_{k-1}\in\mathcal{N}}$ $C_{k-1}^{1}\beta_{ii_1\ldots i_{k-1}}\bar{x}_{i_2}\ldots\bar{x}_{i_{k-1}})$, the inequality $\delta_i>\alpha_{i3}$ holds for all $i\in\mathcal{N}$. Furthermore, the establishment of the above condition and $\bar{x}<\frac{(\sum_{k=3}^{n}\sum_{i_1,\ldots,i_{k-1}\in\mathcal{N}}C_{k-1}^{1}\beta_{ii_1\ldots i_{k-1}}\bar{x}_{i_2}\ldots\bar{x}_{i_{k-1}})}{(\sum_{j\in\mathcal{N}}\beta_{ij}+\sum_{k=3}^{n}\sum_{i_1,\ldots,i_{k-1}\in\mathcal{N}}C_{k\!-\!1}^{1}\beta_{ii_1\ldots i_{k\!-\!1}}\bar{x}_{i_2}\ldots\bar{x}_{i_{k-1}})}$ contribute to promoting the emergence of bistable behavior.
\end{remark}

\section{Numerical examples}

~~~~In this section, we conduct numerical simulations to demonstrate the validity of our analytical findings. 

\subsection{Approximate performance of the mean-field SIS social contagion processes corresponding to the $2^n$-state Markov chain models}

In this subsection, the approximate effectiveness of the mean-field SIS social contagion processes on weighted and directed hypergraphs is observed by comparing it to the Markov Chain model with $2^n$ states. A series of simulations are conducted on two different weighted and directed hypergraphs with 102 agents. In both hypergraphs, the dyadic links form a strongly connected Barab{\'a}si–Albert network, and 10,000 directed third-order links are randomly assigned. By randomly selecting two different types of reproduction numbers $\rho(I-h\mathcal{D}+h\mathcal{B})$ associated with network topology and transition rates, the mean-field SIS social contagion processes converge to the healthy state and endemic equilibrium, respectively. The sampling period $h$ and terminal time are set to $0.1$ and $100$, respectively. For brevity's sake, each simulation represents the average infection level of the entire population of agents.
\begin{figure}[!htbp]
	\centering
	\includegraphics[width=0.5\linewidth]{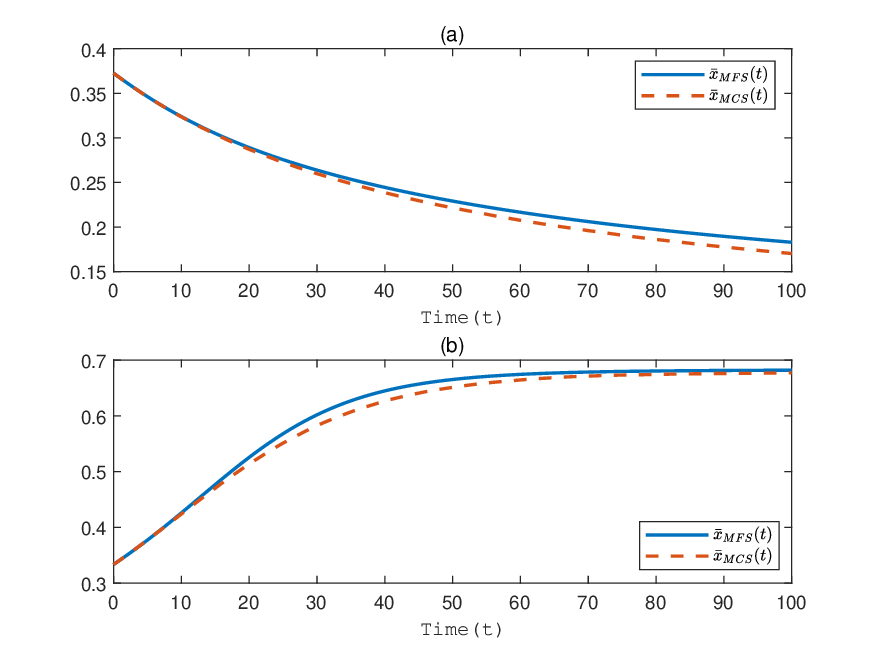}
	\caption{(a) and (b) show the state trajectories of the infection level for the mean-field SIS social contagion processes and Markov chain model when the reproduction number $\rho(I-h\mathcal{D}+h\mathcal{B})$ equals 0.9995 and 1.0056, respectively. }
	\label{Fig: compare}
\end{figure}

In addition to the mean-field SIS social contagion processes, another obstacle lies in simulating the Markov chain model with $2^n$ ($n = 102$) states. Given the impracticality of direct simulations due to the sheer volume of states, we turn to Monte Carlo simulations \cite{rubinstein2016simulation}. Employing the previously mentioned parameters and initial conditions, we run Monte Carlo simulations 50,000 times to demonstrate the efficacy of the $2^n$-state Markov chain model. The initial compartments of the nodes are randomly assigned with probabilities of being susceptible and infected, set at $\frac{2}{3}$ and $\frac{1}{3}$, respectively.

Based on the above configurations, the comparison results are plotted in Fig. \ref{Fig: compare}. To show the difference between the mean-field SIS social contagion processes and the Markov chain model with $2^n$ states, these notations $\bar{x}_{MFS}$ and $\bar{x}_{MCS}$ represent the average infection level obtained by the mean-field SIS social contagion processes and the Markov chain model, respectively.  Fig. \ref{Fig: compare}(a)  shows the state trajectories of these kinds of models which approach the healthy state when the reproduction number $\rho(I-h\mathcal{D}+h\mathcal{B})$ equals 0.9995. Specifically, the maximum approximate error of these state trajectories is 0.0615. As is displayed in Fig. \ref{Fig: compare}(b), all state trajectories approach the endemic equilibrium when the reproduction number $\rho(I-h\mathcal{D}+h\mathcal{B})$ equals 1.0056 and its maximum approximate error is 0.0404. It is observed that the approximation accuracy of the mean-field SIS social contagion processes is acceptable. 

\begin{figure}[!htbp]
	\centering
	\includegraphics[width=0.5\linewidth]{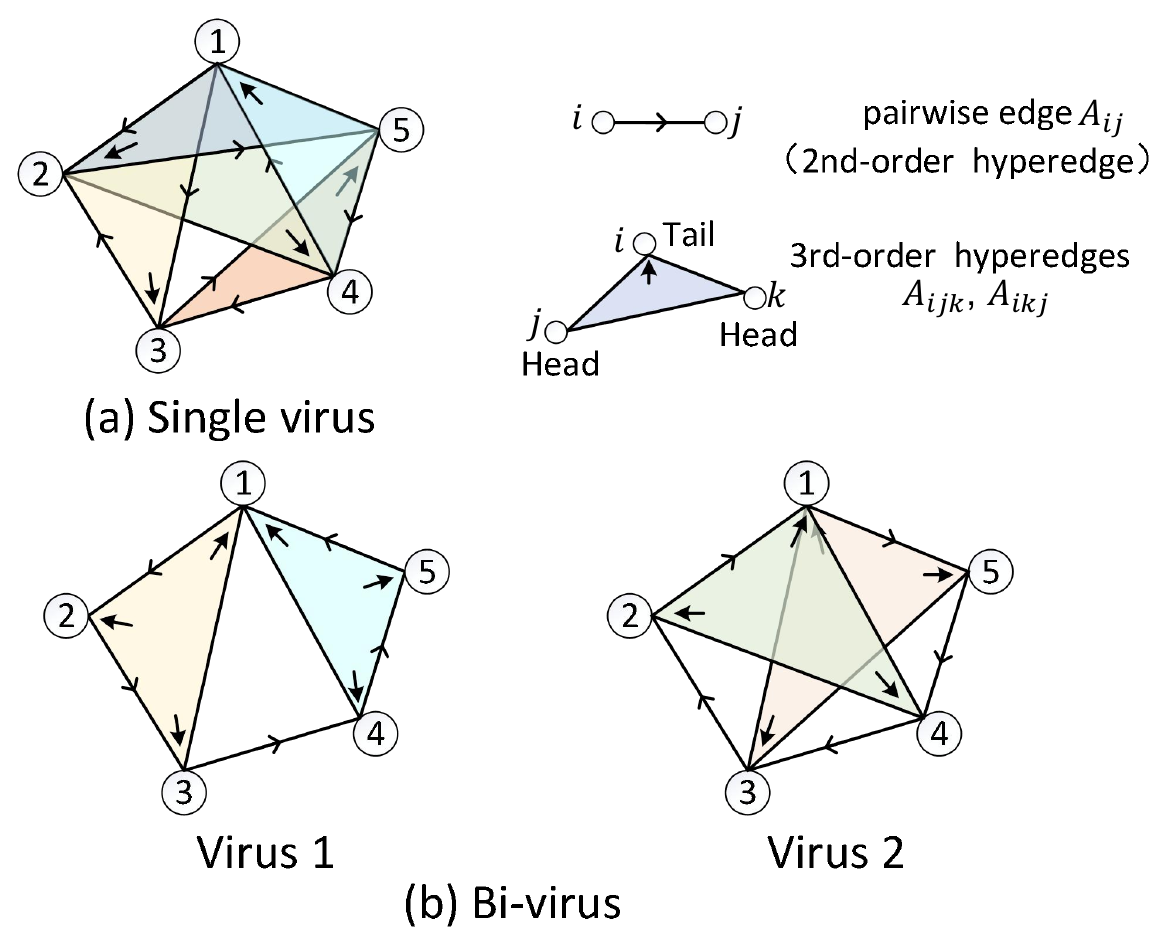}
	\caption{The weighted and directed hypergraph with 2nd-order and 3rd-order hyperedges}
	\label{Fig:hypergraph_example_3}
\end{figure}

\begin{figure}[!htbp]
	\centering
	\subfigure[]{
		\includegraphics[width=0.32\textwidth]{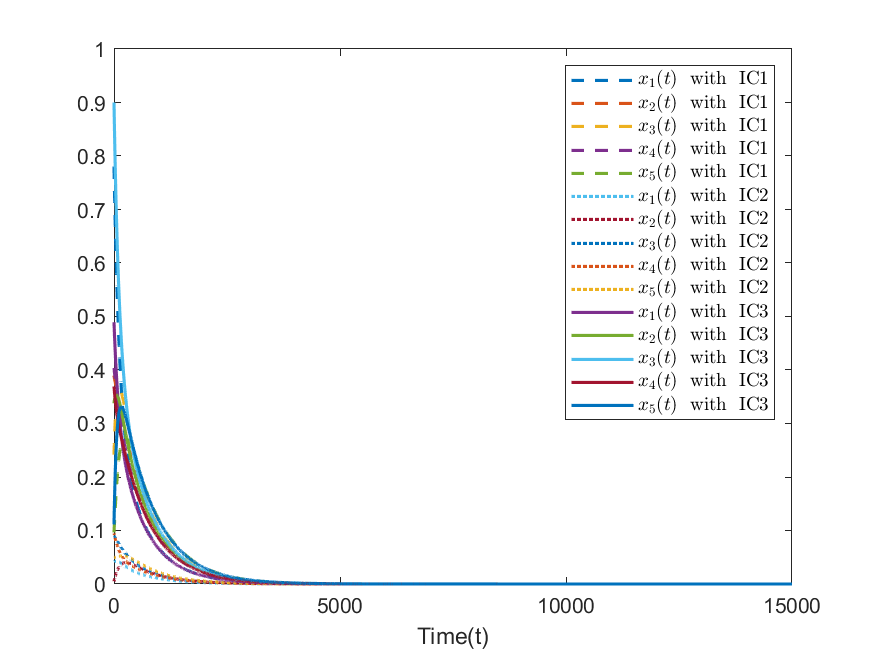}}
	\subfigure[]{
		\includegraphics[width=0.32\textwidth]{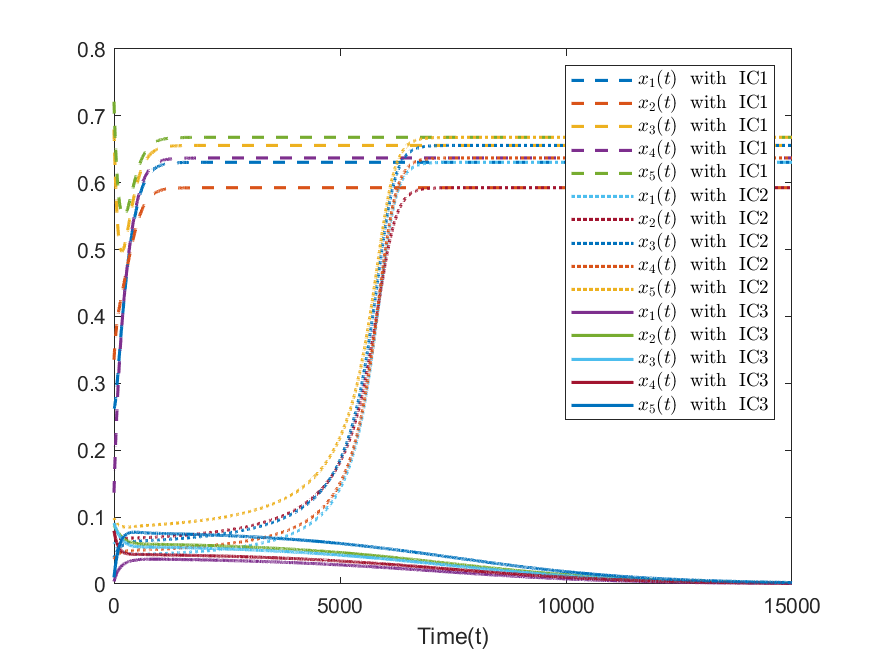}}
	\subfigure[]{
		\includegraphics[width=0.32\textwidth]{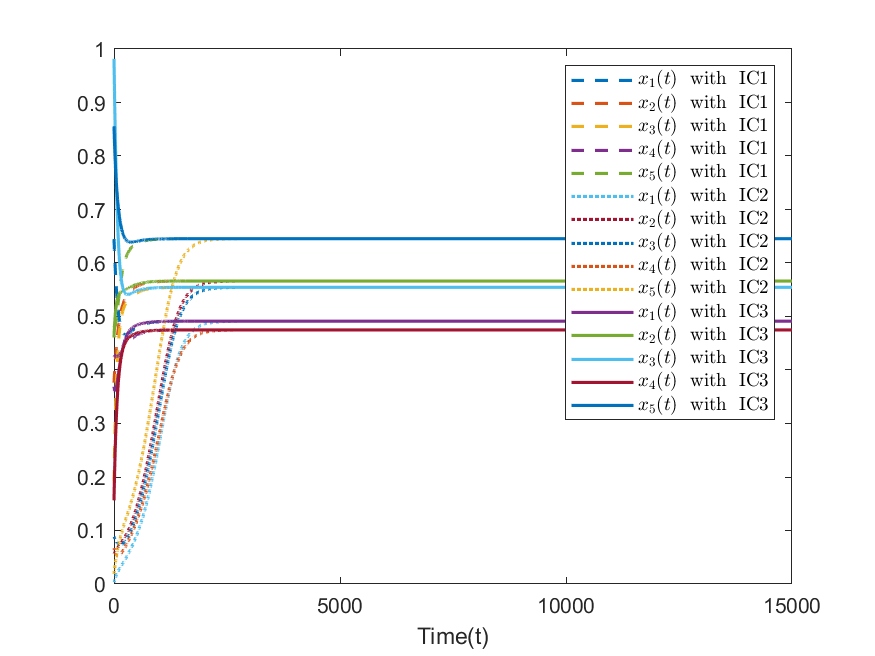}}
	\caption{(a) State trajectory of the infection level $x(t)$ when the reproduction number equals 0.9986. From 3 different random initial conditions, the system (\ref{matrix_form}) always converges to the healthy state. (b) State trajectory of the infection level $x(t)$ when the reproduction number equals 0.9995. From 3 different random initial conditions, the system (\ref{matrix_form}) either converges to the corresponding endemic equilibrium (0.6306,0.5926,0.6559,0.6371,0.6679) or to the healthy state. (c) State trajectory of the infection level $x(t)$ when the reproduction number equals 1.0021. From 3 different random initial conditions, the system (\ref{matrix_form}) always converges to the corresponding endemic equilibrium (0.4913,0.5663,0.5547,0.4750,0.6455). }	\label{Fig: Convergence_Performance}
\end{figure}

\begin{figure}[!htbp]
	\centering
	\includegraphics[width=0.5\linewidth]{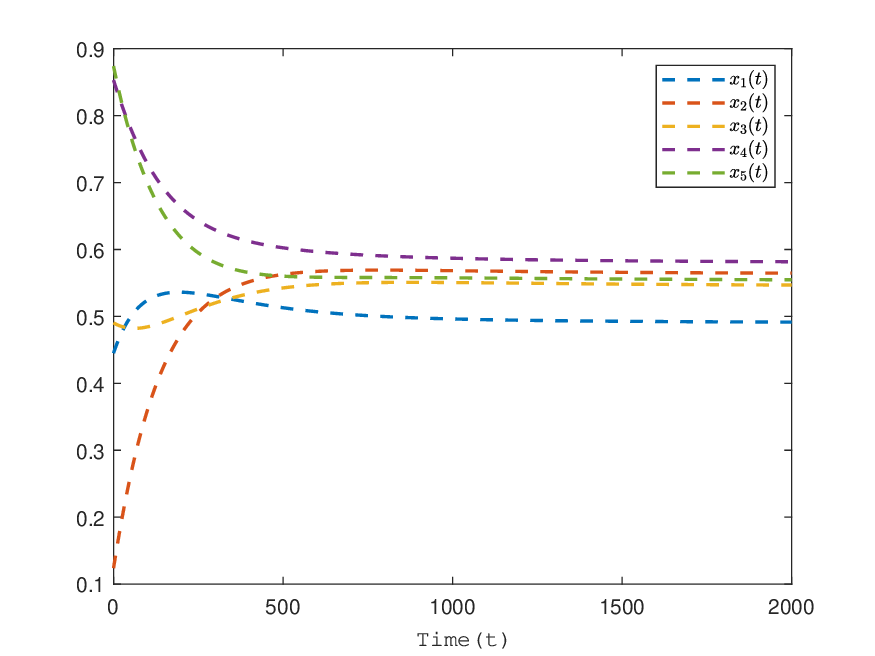}
	\caption{State trajectory of the given simulated infection level $x(t)$ }
	\label{Fig: Parameter_Identification}
\end{figure}

\begin{table}[!htbp]
\begin{center}
\caption{The heterogeneous spread parameters for the simulated system displayed in Fig. \ref{Fig: Parameter_Identification} }
\centering
\begin{tabular}{c| c | c | c | c | c}
\hline
\hline
 Node $i$ & 1 & 2 & 3 & 4 & 5  \\
\hline
 $\delta_i$ & 0.2900  &  0.2500  &  0.2100  &  0.2200 &  0.3500 \\
 \hline
$\mu_i$ &  0.6000  &  0.7300 &   0.5300  &  0.8000 &   0.6500  \\
\hline
$\mu_{i3}$ &    0.6000  &  0.7000   & 0.4000  &  0.8000  &  0.5700   \\
\hline
\hline
\end{tabular}
\end{center}
\end{table}

In what follows, we present a simulation that implements the parameter estimation algorithm discussed in Section 4 to evaluate their performance with clean data. Although the data utilized in this subsection are synthetic, the insights obtained from these exercises contribute significantly to understanding our approach. Consider the system (\ref{3_agents-hyperedge_SIS_model}) with the weighted and directed hypergraph presented in Fig. \ref{Fig:hypergraph_example_3} and the following parameters as $n=5, T=2000, h=0.01$. The weight distribution of the hyperedges in this hypergraph is presented as follows: 
$A_{14}=0.4896, A_{21}= 0.1925, A_{23}=0.1231, A_{31}=0.2055, A_{34}=0.1465, A_{45}=0.1891, A_{52}=0.0427, A_{53}=0.6352, A_{145}=A_{154}=0.2819, A_{215}=A_{251}=0.5386, A_{312}=A_{321}=0.6952, A_{412}=A_{421}=0.4991, A_{534}=A_{543}=0.5358$.

The optimal parameters $\theta_i^{*}=[\delta_i \ \mu_{i} \ \mu_{i3}]^{\top}$ are derived through the utilization of the \textit{lsqlin} function within Matlab, employing an interior-point algorithm. Specifically, for the given state $x(t)$ shown in Fig. \ref{Fig: Parameter_Identification}, the learned spread parameters are obtained in Table 1 in light of Problem \ref{Parameter_Identification_th}. 

\subsection{Convergence Performance of the Discrete-Time SIS Social Contagion Processes}
\subsubsection{Conventional higher-order single SIS Social Contagion Process}
In this subsection, we concentrate on the convergence performance of the single SIS social contagion processes over weighted and directed hypergraph presented in Fig. \ref{Fig:hypergraph_example_3}a. Each of these hyperedges possesses a weight of 1. Significantly, the underlying graph of this hypergraph, which solely represents pairwise associations, exhibits strong connectivity.
The sampling period $h$ is set to $0.01$. We choose three sets of configurations with regard to the transition rates $\delta_i, \mu_i$ and $\mu_{i3}$: first, $\mathcal{D}=\mathsf{diag}([0.6 \ \ 0.65 \ \ 0.6 \ \ 0.56 \ \ 0.65]^{\top}),$ $\mathcal{B}^2=\mathsf{diag}([\mu_{1} \ \ \mu_{2} \ \ \ldots \ \  \mu_{5}]^{\top})=\mathsf{diag}([0.31 \ \ 0.30 \ \ 0.32 \ \ 0.29 \ \ 0.33]^{\top}),$ and $\mathcal{B}^3=\mathsf{diag}([\mu_{31} \ \ \mu_{32} \ \ \ldots \ \  \mu_{35}]^{\top})=\mathsf{diag}([0.25 \ \ 0.31 \ \ 0.34 \ \ 0.35 \ \ 0.26]^{\top}),$ second, $\mathcal{D}=\mathsf{diag}([0.49 \ \ 0.5 \ \ 0.51 \ \ 0.52 \ \ 0.5]^{\top}),$ $\mathcal{B}^2=\mathsf{diag}([0.3183 \ \ 0.3173 \ \ 0.3253 \ \ 0.2543 \ \ 0.3263]^{\top}),$ and $\mathcal{B}^3=\mathsf{diag}([0.7447 \ \ 0.7578 \ \ 0.7487 \ \ 0.9940 \ \ 0.7157]^{\top}),$ third, $\mathcal{D}=\mathsf{diag}([0.6 \ \ 0.5 \ \ 0.6 \ \ 0.7 \ \ 0.5]^{\top}),$ $\mathcal{B}^2=\mathsf{diag}([0.51 \ \ 0.50 \ \ 0.52 \ \ 0.55 \ \ 0.53]^{\top}),$ and $\mathcal{B}^3=\mathsf{diag}([0.55 \ \ 0.41 \ \ 0.44 \ \ 0.5 \ \ 0.6]^{\top})$. With these configurations, Assumptions 2-3 are already satisfied, and their reproduction number $\rho(I-h\mathcal{D}+h\mathcal{B})$ equal 0.9986, 0.9995, and 1.0021, respectively.

In each figure, the abbreviation "IC" represents the initial condition. As for three sets of different random IC, we perform simulations on the system (\ref{matrix_form}), and the results depicted in Fig. \ref{Fig: Convergence_Performance} align with the analytical outcomes presented in this paper. Notably, we observe bi-stability in the dynamics of the system (\ref{matrix_form}).

\subsubsection{Bi-virus SIS Social Contagion Processes over hypergraphs}
\begin{figure}[!htbp]
	\centering
	\subfigure[]{
		\includegraphics[width=0.32\textwidth]{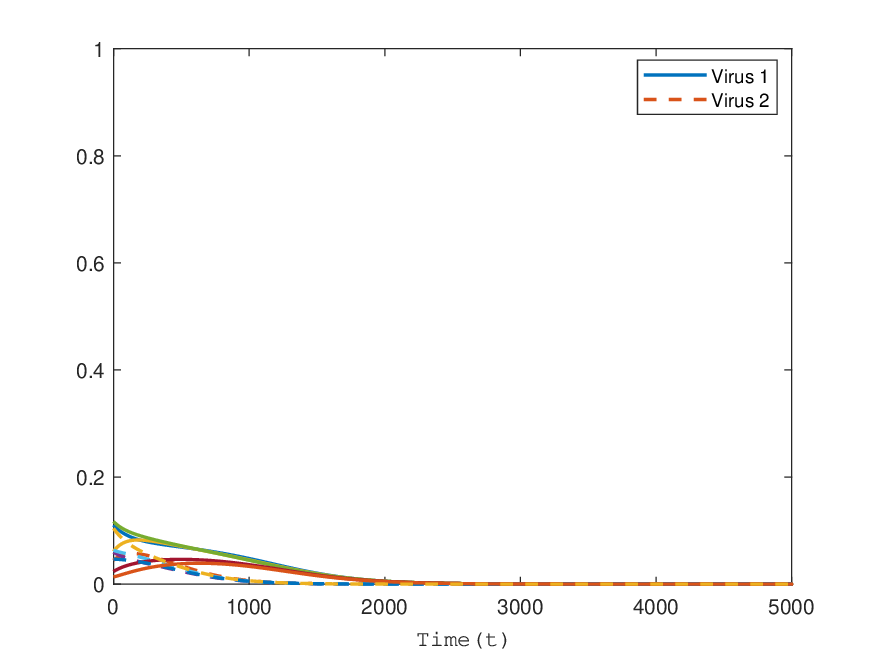}}
	\subfigure[]{
		\includegraphics[width=0.32\textwidth]{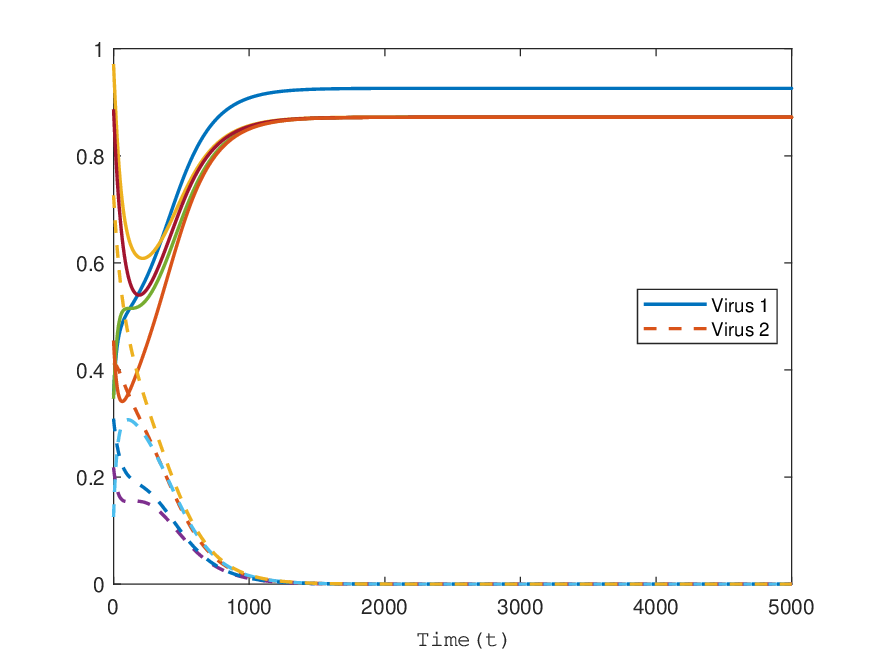}}
	\subfigure[]{
		\includegraphics[width=0.32\textwidth]{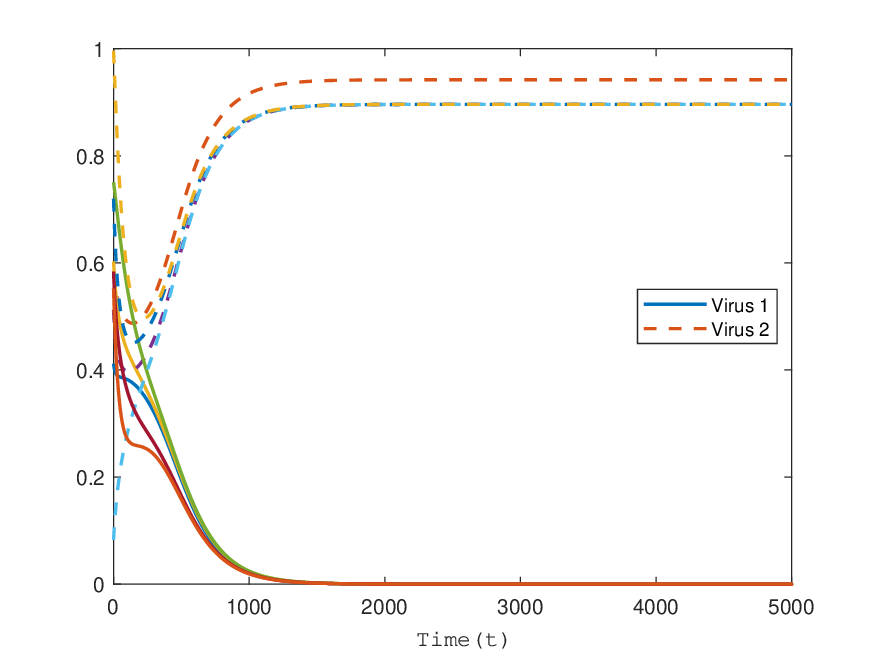}}
	\caption{ State trajectory  $x_{[\ell]}(t)(\ell=1,2)$ of the bi-virus system (\ref{bi_virus_2}) for different random initial conditions when the reproduction numbers equal 0.9970 and 0.9960. }\label{bivirus_fig1}
\end{figure}

\begin{figure}[!htbp]
	\centering
	\subfigure[]{
		\includegraphics[width=0.4\textwidth]{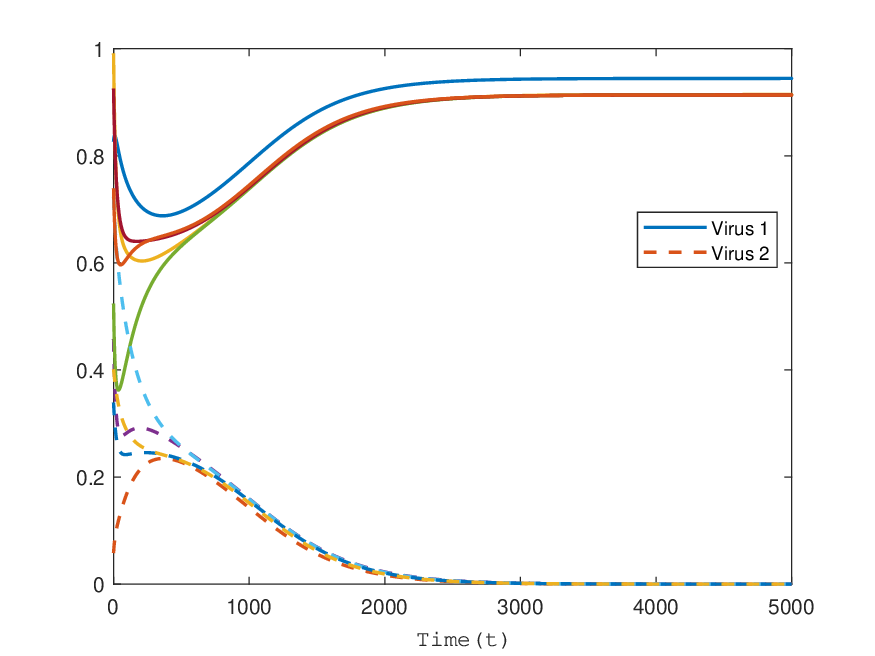}}
	\subfigure[]{
		\includegraphics[width=0.4\textwidth]{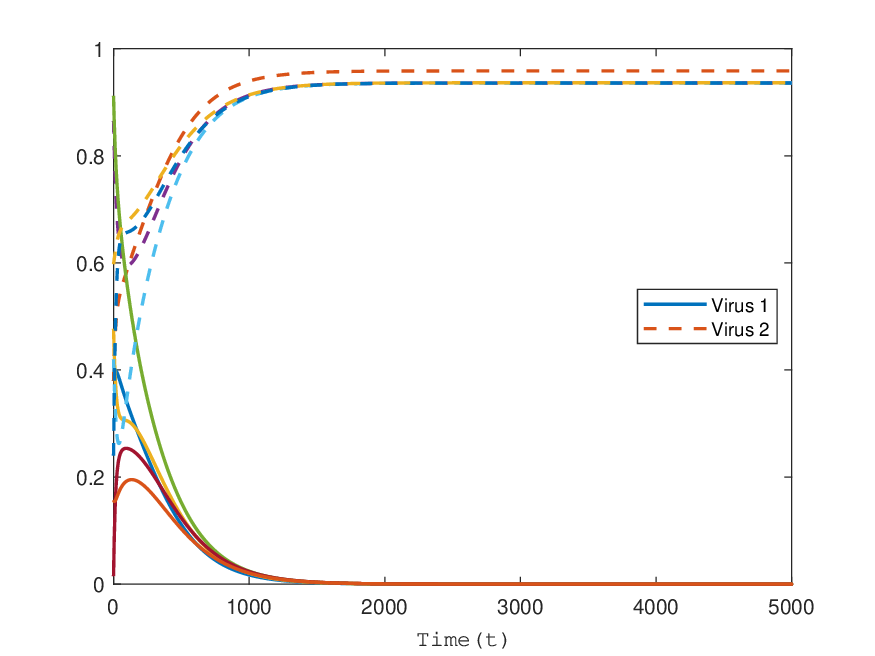}}
	
	\caption{ State trajectory  $x_{[\ell]}(t)(\ell=1,2)$ of the bi-virus system (\ref{bi_virus_2}) for different random initial conditions when the reproduction numbers equal 1.0150 and 1.0250. }\label{bivirus_fig2}
\end{figure}

The bi-virus  SIS Social Contagion Processes characterized by a hypergraph as depicted in Fig. 2b are under consideration. The hypergraph has $n=5$ nodes and the weights of all its hyperedges are equal to 1. It's pairwise
interactions are captured by two-cycle graphs, which implies that Assumption \ref{bi_virus_assumption2} holds. In our simulations, $x_{\ell i}(0)$ is initially sampled from a uniform distribution over the interval $(0, 1)$. Subsequently, the vectors $x_{[1]}(0)$ and $x_{[2]}(0)$ are normalized to guarantee that Assumption \ref{bi_virus_assumption1} holds. Let $\mathcal{D}_1=\mathcal{D}_2=0.5I,h = 0.01.$ The intrinsic parameters are given as $\mu_{1i}=0.2,\mu_{1i3}=2,\mu_{2i}=0.1,\mu_{2i3}=2.5.$ With these configurations, the inequalities of both conditions for Proposition \ref{bi_virus_th2} are already satisfied, and their reproduction numbers $\rho(I-h\mathcal{D}_{\ell}+h\mathcal{B}_{\ell})$ equal 0.9970 and 0.9960, respectively. As observed from Fig. \ref{bivirus_fig1}, for different random initial conditions, the system (\ref{bi_virus_2}) converges either to the healthy state $(\textbf{0}_n,\textbf{0}_n)$ or to the dominant endemic equilibrium $(\bar{x}_{[1]},\textbf{0}_n)$ or$(\textbf{0}_n,\bar{x}_{[2]})$. In other words, both the dominant endemic equilibria $(\bar{x}_{[1]},\textbf{0}_n)$ or$(\textbf{0}_n,\bar{x}_{[2]})$  are simultaneously locally asymptotically
stable, which is impossible for a bi-virus system without high-order interactions. Additionally, we consider $\mu_{1i}=2,\mu_{1i3}=2,\mu_{2i}=3,\mu_{2i3}=2.5,$ which yields that the reproduction numbers $\rho(I-h\mathcal{D}_{\ell}+h\mathcal{B}_{\ell})$ equal 1.0150 and 1.0250, respectively. Under this case, for different initial conditions, there are two
locally asymptotically stable two dominant endemic equilibria $(\bar{x}_{[1]},\textbf{0}_n)$ and $(\textbf{0}_n,\bar{x}_{[2]})$ in Fig. \ref{bivirus_fig2}. However, no
trajectories in $\mathcal{S}$ converge the healthy state $(\textbf{0}_n,\textbf{0}_n)$ except if one starts at 
$(\textbf{0}_n,\textbf{0}_n)$.

\section{Conclusions}
This article presents a rough bifurcation analysis of the discrete-time SIS social contagion processes on a weighted and directed hypergraph. Our investigation primarily centers on examining the system’s dynamics in the healthy state and the endemic behaviors. Notably, we observe that due to higher-order interactions, bistability can arise even when the reproduction number is less than 1. Moreover, when the reproduction number exceeds 1, the unique endemic equilibrium is globally stable provided that the higher-order terms remain sufficiently small. Note that we also provide some novel conditions of the healthy state and the endemic behaviors for the system and the corresponding domains of attraction with the help of tensor algebra. It is shown that the system is strongly monotone. Finally, we provide several numerical examples to validate the theoretical findings. More specially, the effectiveness of the SIS model as an approximation of the Markov chain model is displayed. The example shows the performance of the parameter learning algorithm and the convergence performance of the discrete-time SIS model is demonstrated.

\bibliographystyle{ieeetr}
\bibliography{reference}

\section{Appendix}
\subsection{Proof of Proposition 1}
It is obvious that a healthy state always exists. In light of Assumption 2, we can obtain that $\mathcal{D}^{-1}\mathcal{B}+\mathcal{D}^{-1}\mathcal{H}z$ is an irreducible non-negative matrix. By applying Perron-Frobenius theorem \cite{horn2012matrix}, there exist a strictly positive left Perron-Frobenius eigenvector $\xi$ and its associated eigenvalue $\upsilon=\rho(\mathcal{D}^{-1}\mathcal{B}+\mathcal{D}^{-1}\mathcal{H}z)$ such that $\xi^{\top}(\mathcal{D}^{-1}\mathcal{B}+\mathcal{D}^{-1}\mathcal{H}z)=\upsilon\xi^{\top}.$ Construct the Lyapunov function $V_1(t)=\xi^{\top}\mathcal{D}^{-1}x(t)$. Clearly, $V_1(t) \geq 0, \forall t\geq 0$ and $V_1(t)=0$ if and only if $x(t)=\textbf{0}_n$. It follows that
\begin{align}
\Delta V_1(t) 
&=h\xi^{\top}\mathcal{D}^{-1}\left(-\mathcal{D}x(t)+(I-\mathsf{diag}(x(t)))(\mathcal{B}x(t) +\mathcal{H}x^2(t))\right) \notag\\
&\leq h\xi^{\top}\mathcal{D}^{-1}\left(-\mathcal{D}x(t)+\mathcal{B}x(t) +\mathcal{H}x^2(t)\right) \notag\\
&\leq h\xi^{\top}\mathcal{D}^{-1}\left(-\mathcal{D}x(t)+(\mathcal{B} +\mathcal{H}z)x(t)\right) \notag\\
&=-h\xi^{\top}\mathcal{D}^{-1}\mathcal{D}x(t)+h\upsilon\xi^{\top}x(t)\notag\\
&=h(-1+\upsilon)\xi^{\top}\mathcal{D}^{-1}\mathcal{D}x(t)\notag\\
&<h(-1+\upsilon)\mbox{min}_{i\in\mathcal{N}}\{\delta_i\}V_1(t)\notag\\
&=-\alpha V_1(t)
\end{align}
where $\alpha=h(1-\upsilon)\mbox{min}_{i\in\mathcal{N}}\{\delta_i\}.$ According to the Assumption 3a and $\upsilon<1$, one gets $\alpha\in(0,1).$ Therefore, one can derive that $V_1(t+1)\leq(1-\alpha)V_1(t)$, which recursively gives $V_1(t)\leq(1-\alpha)^tV_1(0)=e^{-\eta t}V_1(0)$ with $\eta=-\mbox{ln}(1-\alpha)>0.$ Based on the define of $V_1(t)$, $\mbox{min}_{i\in\mathcal{N}}\{\delta_{i}^{-1}\}\xi_ix_i(t)\leq \mbox{max}_{i\in\mathcal{N}}\{\delta_{i}^{-1}\}\sum_{i\in\mathcal{N}}\xi_ix_i(0)e^{-\eta t},$ which implies $x_i(t)\leq \frac{\mbox{min}_{i\in\mathcal{N}}\{\delta_{i}\}}{\mbox{max}_{i\in\mathcal{N}}\{\delta_{i}\}} \frac{\sum_{i\in\mathcal{N}}\xi_ix_i(0)e^{-\eta t}}{\xi_i}\leq\frac{\xi^{\top}x(0)}{\xi_i}e^{-\eta t}.$ This indicates that the healthy state is exponentially stable.

In what follows, employ the method of proof by contradiction to establish its globality. Assuming there exist an equilibrium $\bar{x}$ different than the origin and meets $\bar{x}=\bar{x}+h\left(-\mathcal{D}\bar{x}+(I-\mathsf{diag}(\bar{x}))(\mathcal{B}\bar{x} +\mathcal{H}\bar{x}^2)\right)$. Define the map $\mathcal{T}(x(t))=\mathscr{T}(\mathcal{D}^{-1}(\mathcal{B}x(t)+\mathcal{H}x^2(t))) =[\mathscr{T}_1(x(t)),\mathscr{T}_2(x(t)),\ldots,\mathscr{T}_n(x(t))]^{\top}$. For all $i\in\mathcal{N},$ $\mathscr{T}_i(x(t))=\frac{(\mathcal{D}^{-1}(\mathcal{B}x(t)+\mathcal{H}x^2(t)))_i}{1 +(\mathcal{D}^{-1}(\mathcal{B}x(t)+\mathcal{H}x^2(t)))_i}$. In other words, $\bar{x}$ is the fixed point of the map $\mathcal{T}(x(t)).$ 
Moreover, since $\rho(\mathcal{D}^{-1}\mathcal{B}+\mathcal{D}^{-1}\mathcal{H}z)<1,$   $\lim_{k\rightarrow\infty}\mathcal{T}^k(x(t))=\textbf{0}_{n}$ by employing the fact (i) in Theorem 5.1 of \cite{cisneros2021multigroup}. On the other hand, since $\bar{x}$ is a fixed point of the map $\mathcal{T}(x(t))$, $\mathcal{T}^k(\bar{x})=\bar{x}$. Then, $\bar{x}=\textbf{0}_{n},$ which is contradictory to the assumption. Finally, this shows that the healthy state is globally exponentially stable. 
  {\hfill$\square$}
  
\subsection{Proof of Proposition 2}
The Jacobian matrix at the origin of (\ref{matrix_form}) is $J(\textbf{0}_n)=I-h\mathcal{D}+h\mathcal{B}.$ According to Theorem 42 in Section 5.9 of the work \cite{vidyasagar2002nonlinear}, since $\rho(I-h\mathcal{D}+h\mathcal{B})<1,$ the healthy state is locally asymptotically stable.

Demonstrating the existence of an endemic equilibrium point in the system (\ref{matrix_form}) involves establishing the presence of a fixed point within the mapping function $\mathcal{T}(x(t))=\mathscr{T}(\mathcal{D}^{-1}(\mathcal{B}x(t)+\mathcal{H}x^2(t)))$. Define set $\Omega=\{\omega\in[0,1]^n|\frac{1}{2}z_i\leq\omega_i\leq1,i\in\mathcal{N}\}.$  Building upon the demonstration of fact (iv) in Theorem 5.1 as presented in \cite{cisneros2021multigroup} and recalling that $\theta\geq4,$ there exists a $\bar{x}\in\Omega$ such that $\mathcal{T}(\bar{x})=\bar{x},$ i.e. an endemic equilibrium point $\bar{x}\geq\frac{1}{2}\textbf{1}_n$ for the system (\ref{matrix_form}), which belongs to $\Omega.$

 The Jacobian matrix at endemic equilibrium point $\bar{x}$ can be calculated as $J(\bar{x})=I+hJ_1(\bar{x}),$ where $J_1(\bar{x})=-\mathcal{D}+(I-\mathsf{diag}(\bar{x}))(\mathcal{B}+2\mathcal{H}\bar{x}) -\mathsf{diag}(\mathcal{B}\bar{x}+\mathcal{H}\bar{x}^2).$ Combining Lemma 1 and Assumption 2, $J_1(\bar{x})$ is a Metzler matrix.  Besides, since $1-2\bar{x}\leq0$, $J_1(\bar{x})$ is a Hurwitz matrix based on the fact (v) in Theorem 5.1 as presented in \cite{cisneros2021multigroup}. In other words, the following inequality holds: $s(J_1(\bar{x}))<0.$

On the other hand, following Assumptions 2 and 3b, $J(\bar{x})=I+hJ_1(\bar{x})$ is a nonnegative irreducible matrix. Based on the Perron Frobenius Theorem, one gains that
there exists $\vartheta\gg0,$ such that
\begin{align*}
(I+hJ_1(\bar{x}))\vartheta=\rho(I+hJ_1(\bar{x}))\vartheta.
\end{align*}
Then,
\begin{align*}
hJ_1(\bar{x})\vartheta=\rho(hJ_1(\bar{x}))\vartheta=s(hJ_1(\bar{x}))\vartheta,
\end{align*}
which implies
\begin{align*}
hs(J_1(\bar{x}))<0\Leftrightarrow\rho(I+hJ_1(\bar{x}))<1.
\end{align*}
Therefore, the endemic equilibrium point $\bar{x}$ is locally asymptotically stable.  {\hfill$\square$}

\subsection{Proof of Proposition 3}
Since $\rho(J(\textbf{0}_n))= \rho(I-h\mathcal{D}+h\mathcal{B})>1$, the healthy state is unstable. According to Assumptions 2 and 3b and by employing the Perron Frobenius Theorem for nonnegative irreducible matrices, $\rho(I-h\mathcal{D}+h\mathcal{B})= s(I-h\mathcal{D}+h\mathcal{B})$. Since $\rho(I-h\mathcal{D}+h\mathcal{B})>1$, there exists $\varphi\gg0$ such that $(I-h\mathcal{D}+h\mathcal{B})\varphi= \rho(I-h\mathcal{D}+h\mathcal{B})\varphi>\varphi$. Therefore, one gets that $(-h\mathcal{D}+h\mathcal{B})\varphi= \rho(-h\mathcal{D}+h\mathcal{B})\varphi=s(-h\mathcal{D}+h\mathcal{B})\varphi>0\varphi$, which yields $\rho(I-h\mathcal{D}+h\mathcal{B})>1\Leftrightarrow hs(-\mathcal{D}+\mathcal{B})>0$. Then, one has $\rho(\mathcal{D}^{-1}\mathcal{B})>1.$ Therefore, based on the fact (vii) in Theorem 5.1 of \cite{cisneros2021multigroup}, there exists an endemic equilibrium point $\bar{x}\gg\textbf{0}_n$ for the system (\ref{matrix_form}).

The system (\ref{matrix_form}) can be reestablished as
\begin{align}\label{matrix_form1}
 \tilde{x}(t+1)=(I+h\mathcal{G}(x(t),\bar{x}))\tilde{x}(t),
\end{align}
where $\tilde{x}(t)=x(t)-\bar{x}$ and $\mathcal{G}(x(t),\bar{x})=-\mathcal{D}+(I-\mathsf{diag}(\bar{x}))\mathcal{B} -\mathsf{diag}(\mathcal{B}x(t))+(I-\mathsf{diag}(\bar{x}))[(x(t)+\bar{x})^{\top} \frac{H_1+H_1^{\top}}{2} \ \ (x(t)+\bar{x})^{\top} \frac{H_2+H_2^{\top}}{2} \ \ \ldots \ \ (x(t)+\bar{x})^{\top} \frac{H_n+H_n^{\top}}{2} ]^{\top}-\mathsf{diag}(\mathcal{H}x^2(t))$ with $H_i=\left[\beta_{ijk}\right]_{n\times n}$ for all $i\in\mathcal{N}$.
In addition, let $\zeta=\bar{x}$, one can attain that for all $i\in\mathcal{N}$
\begin{align*}
  (h\mathcal{G}(x(t),\bar{x})\zeta)_i=&h\left(-\delta_i\zeta_i+(1-\bar{x}_i)\sum_{j\in\mathcal{N}} \beta_{ij}\zeta_j-\sum_{j\in\mathcal{N}}\beta_{ij}x_j(t)\zeta_i+(1-\bar{x}_i) \sum_{j,k\in\mathcal{N}}(x_j(t)+\bar{x}_j)\frac{\beta_{ijk} +\beta_{ikj}}{2}\zeta_k\right.\\
   &  \ \ \ \ \ \ \ \ \left.-\zeta_i\sum_{j,k\in\mathcal{N}}\beta_{ijk}x_j(t)x_k(t)\right)\\
  =&h\left(-\sum_{j\in\mathcal{N}}\beta_{ij}x_j(t)\zeta_i +(1-\bar{x}_i) \sum_{j,k\in\mathcal{N}}x_j(t)\frac{\beta_{ijk} +\beta_{ikj}}{2}\zeta_k--\zeta_i\sum_{j,k\in\mathcal{N}}\beta_{ijk}x_j(t)x_k(t) \right)\\
  \leq&h\left(-\sum_{j\in\mathcal{N}}\beta_{ij}c_j\zeta_i +(1-\bar{x}_i) \sum_{j,k\in\mathcal{N}}\frac{\beta_{ijk} +\beta_{ikj}}{2}\zeta_k-\zeta_i\sum_{j,k\in\mathcal{N}}\beta_{ijk}c_jc_k \right)\\
  \leq&h\left(-\sum_{j\in\mathcal{N}}\beta_{ij}c_j\zeta_i +(1-\bar{x}_i) \sum_{j,k\in\mathcal{N}}\frac{\beta_{ijk} +\beta_{ikj}}{2}\zeta_k \right)\\
\end{align*}
 Based on the Assumptions 2 and 3b and $c\geq\textbf{0}_{n}$, one can observe that for $i,j,k\in\mathcal{N},$ if the $\beta_{ijk}$ is sufficiently small, then $(h\mathcal{G}(x(t),\bar{x})\zeta)_i\leq-\varpi\zeta_i,$ for some constant $\varpi\in(0,1).$  Notice that $I+h\mathcal{G}(x(t),\bar{x})$ is a nonnegative matrix, since the Assumptions 2 and 3b and Lemma 1. Then, one has
 \begin{align*}
   &\|I+h\mathcal{G}(x(t),\bar{x})\|_{\infty,\mathsf{diag}\{\zeta\}^{-1}} \\ =&\|I+h\mathsf{diag}\{\zeta\}^{-1}\mathcal{G}(x(t),\bar{x})\mathsf{diag}\{\zeta\}\|_{\infty}\\ =&\mbox{max}_{i\in\mathcal{N}}\left(|1+h\zeta_i^{-1}(\mathcal{G}(x(t),\bar{x}))_{ii} \zeta_i|+\sum_{j\neq i,j\in\mathcal{N}}|h\zeta_i^{-1}(\mathcal{G}(x(t),\bar{x}))_{ij} \zeta_j|\right)\\
   =&\mbox{max}_{i\in\mathcal{N}}\left(1+h\zeta_i^{-1} \sum_{j\in\mathcal{N}}(\mathcal{G}(x(t),\bar{x}))_{ij} \zeta_j\right)\\
   \leq&1-\varpi.
 \end{align*}
Suppose that $V_2(t)=\|x(t)-\bar{x}\|$ is a Lyapunov functional for the system (\ref{matrix_form}). In what follows, calculating the time-derivative of $V_2(t)$, it follows from (\ref{matrix_form1}) that
\begin{align*}
  \Delta V_2(t)=V_2(t+1)-V_2(t) =&\|x(t+1)-\bar{x}\|-\|x(t)-\bar{x}\| \\
  \leq&(\|I+h\mathcal{G}(x(t),\bar{x})\|-1)\|x(t)-\bar{x}\|\\
  \leq&-\varpi \|x(t)-\bar{x}\|\\
  =&-\varpi V_2(t)
\end{align*}
The following inequality can be derived through step-by-step deduction: $V_2(t)\leq(1-\varpi)^tV_2(0),$ which yields $V_2(t)\leq\mbox{e}^{-\kappa t}V(0)$, where $\kappa=-\mbox{ln}(1-\varpi)$. Therefore, the endemic equilibrium point $\bar{x}$ is exponentially stable.    

On the other hand, due to the fact that $\widetilde{\Omega}$ is compact, for each $\varsigma$, the set $\{x\in\widetilde{\Omega}|V(t)\leq\varsigma\}$ is compact, which yields that the Lyapunov function is globally proper. Subsequently, considering the facts that $V(t)$ is positive definite on $\widetilde{\Omega}$ and strictly decreasing for $x(t)\neq \bar{x}$ on $\widetilde{\Omega}$, since $\Delta V_2(t)<0$. Therefore, the endemic equilibrium point $\bar{x}$ is the unique equilibrium, which is globally exponentially stable.   {\hfill$\square$}

\subsection{Proof of Proposition 4}
The dynamics of each virus (\ref{bi_virus_2}) are compared with their single virus counterparts (\ref{matrix_form}) under the same system parameters. It is evident that the dynamics of each virus (\ref{bi_virus_2}) are upper bounded by the dynamics of their corresponding single virus counterparts (\ref{matrix_form}). Then, one completes the proof based on the Proposition \ref{th1} and the comparison principle. {\hfill$\square$}

\subsection{Proof of Proposition 5}
Since the dynamics of each virus (\ref{bi_virus_2}) are upper bounded by the dynamics of their corresponding single virus counterparts (\ref{matrix_form}), then, one completes the proof based on the Theorem \ref{tensor_th1} and the comparison principle. {\hfill$\square$}

\subsection{Proof of Proposition 6}
The Jacobian matrix of the system (\ref{bi_virus_2}) at the healthy state is $J(\textbf{0}_n,\textbf{0}_n)=
\left[
  \begin{array}{cc}
    I-h\mathcal{D}_{1}+h\mathcal{B}_{1} & \textbf{0}_{n\times n} \\
    \textbf{0}_{n\times n} & I-h\mathcal{D}_{2}+h\mathcal{B}_{2} \\
  \end{array}
\right].$ Then, since for each $\ell=1,2,$ $\rho(I-h\mathcal{D}_{\ell}+h\mathcal{B}_{\ell})<1$, $\rho(J(\textbf{0}_n,\textbf{0}_n))<1$, which implies that the
healthy state $(\textbf{0}_n,\textbf{0}_n)$ is locally asymptotically stable. By recalling the Proposition \ref{th2} and $\theta_{\ell}\geq 4$ for each $\ell=1,2$, there exists single-virus endemic equilibrium point $\bar{x}_{[\ell]}\gg\textbf{0}_n$ corresponding to virus $\ell$ such that $\bar{x}_{\ell i}\geq\frac{1}{2}$ for any $i\in\mathcal{N}$ such that $H_{\ell i}\neq\textbf{0}_{n\times n}$, which is locally asymptotically stable. The Jacobian matrix of the system (\ref{bi_virus_2}) at the dominant endemic equilibrium $(\bar{x}_{[1]},\textbf{0}_n)$  is $J(\bar{x}_{[1]},\textbf{0}_n)=
\left[
  \begin{array}{cc}
    I+hJ_{11}(\bar{x}_{[1]},\textbf{0}_n) & J_{12}(\bar{x}_{[1]},\textbf{0}_n) \\
    \textbf{0}_{n\times n} & I+hJ_{22}(\bar{x}_{[1]},\textbf{0}_n) \\
  \end{array}
\right]$ where $J_{11}(\bar{x}_{[1]},\textbf{0}_n)=-\mathcal{D}_{1}+(I-\mathsf{diag}(\bar{x}_{[1]}))(\mathcal{B}_{1}+2\mathcal{H}_{1}\bar{x}_{[1]}) -\mathsf{diag}(\mathcal{B}_{1}\bar{x}_{[1]}+\mathcal{H}_{1}\bar{x}_{[1]}^2)$, $J_{12}(\bar{x}_{[1]},\textbf{0}_n)=-h\mathsf{diag}(\mathcal{B}_{1}\bar{x}_{[1]}+\mathcal{H}_{1}\bar{x}_{[1]}^2),$ and $J_{22}(\bar{x}_{[1]},\textbf{0}_n)=-\mathcal{D}_{2}+(I-\mathsf{diag}(\bar{x}_{[1]}))\mathcal{B}_{2}.$ Combining Proposition \ref{th2}, one gets $\rho(I+hJ_{11}(\bar{x}_{[1]},\textbf{0}_n))<1$. By Assumption, $\rho(I+hJ_{22}(\bar{x}_{[1]},\textbf{0}_n))\leq\rho(I+h(-\mathcal{D}_{2}+\mathcal{B}_{2}))<1$. Then, since $J(\bar{x}_{[1]},\textbf{0}_n)$ is block upper triangular, it is obtained that $\rho(J(\bar{x}_{[1]},\textbf{0}_n))<1$
 which yields that  the dominant endemic equilibrium $(\bar{x}_{[1]},\textbf{0}_n)$ are locally asymptotically stable. By the similar proof,  the dominant endemic equilibrium $(\textbf{0}_n,\bar{x}_{[2]})$ are also locally asymptotically stable.  {\hfill$\square$}

\subsection{Proof of Proposition 7}
Note that for each $\ell=1,2,$ $\rho(I-h\mathcal{D}_{\ell}+h\mathcal{B}_{\ell})>1$, which implies that $\rho(J(\textbf{0}_n,\textbf{0}_n))>1$. Then, the
healthy state $(\textbf{0}_n,\textbf{0}_n)$ is unstable, and there exists a single-virus endemic equilibrium point $\bar{x}_{[\ell]}\gg\textbf{0}_n$ corresponding to virus $\ell$. Since $\rho(I+h(-\mathcal{D}_{\ell}+(I-\textit{diag}\{\bar{x}_{[\imath]}\})\mathcal{B}_{\ell}))>1$ for $\ell,\imath=1,2,\ell\neq\imath,$ there exists at least one coexisting equilibrium $(\hat{x}_{[1]},\hat{x}_{[2]})$ with $\textbf{0}_n\ll\hat{x}_{[1]},\hat{x}_{[2]}\ll\textbf{1}_n$ and $\hat{x}_{[1]}+\hat{x}_{[2]}\ll\textbf{1}_n,$based on Theorem 5.3 in \cite{cui2023general} and Lemma 16 in \cite{cui2024discrete2}, and it is clear that $\rho(J(\bar{x}_{[1]},\textbf{0}_n))>1$. Then, the dominant endemic equilibrium $(\bar{x}_{[1]},\textbf{0}_n)$ are unstable, and the instability of the dominant endemic equilibrium $(\textbf{0}_n,\bar{x}_{[2]})$ can be proved analogously, which complete the proof of i). In what follows, by assumption and based on Proposition \ref{th1} and \ref{th3}, we can get the global stability of the dominant endemic equilibria $(\bar{x}_{[1]},\textbf{0}_n)$ and $(\textbf{0}_n,\bar{x}_{[2]})$.
{\hfill$\square$}

\subsection{Proof of Proposition 9}

The stability proof of the healthy state is the same as in Proposition 2.

 Define set $\widehat{\Omega}=\{\omega\in[0,1]^n|\frac{n-2}{n-1}\tilde{z}_i\leq\omega_i\leq1,i\in\mathcal{N}\}.$ Recalling the definition of $\mathcal{T}(x(t)),$ it is easy to see that $\mathscr{T}(\mathcal{D}^{-1}(\mathcal{B}x(t)+\sum_{k=3}^{n}\mathcal{F}_kx^{k-1}(t))$ is monotonically increasing in $x(t)\geq0$. Then, owing to the fact that $\tilde{\theta}\geq n-1$, for $\omega\in\widehat{\Omega},$
$\mathcal{T}(\omega)=\mathscr{T}(\mathcal{D}^{-1}(\mathcal{B}\omega +\sum_{k=3}^{n}\mathcal{F}_k\omega^{k-1}) 
   \geq  \mathscr{T}(\mathcal{D}^{-1}(\frac{n-2}{n-1}\mathcal{B}\tilde{z} +\sum_{k=3}^{n}(\frac{n-2}{n-1})^{k-1}\mathcal{F}_k\tilde{z}^{k-1} 
   \geq  \mathscr{T}(\frac{n-2}{n-1}\tilde{\theta}\tilde{z}) 
   \geq  \frac{n-2}{n-1}\tilde{z}.
$
 Furthermore, $\mathcal{T}(\omega)=\mathscr{T}(\mathcal{D}^{-1}(\mathcal{B}\omega +\sum_{k=3}^{n}\mathcal{F}_k\omega^{k-1}))\leq\textbf{1}_n.$ Therefore, we can obtain that the map $\mathcal{T}(x(t))$ maps the set $\widehat{\Omega}$ into itself. Based on the Brouwer fixed-point theorem \cite{shapiro2016fixed}, there exists $\bar{x}\in\widehat{\Omega}$ such that $\mathcal{T}(\bar{x})=\bar{x},$ i.e. an endemic equilibrium point $\bar{x}$ for the system (\ref{matrix_form}), which belongs to $\widehat{\Omega}.$

The Jacobian matrix at endemic equilibrium point $\bar{x}$ can be calculated as $\widetilde{J}(\bar{x})=I+h\widetilde{J}_1(\bar{x}),$ where 

\begin{align*} 
(\widetilde{J}_1(\bar{x}))_{ij}=\left\{
  \begin{array}{ll}
-\delta_i-(\mathcal{B}x(t)+\sum_{k=3}^{n}\mathcal{F}_kx^{k-1}(t))_i+(\widetilde{\mathscr{J}}_1(\bar{x}))_{ii}, \ \ \ j=i\\
(\widetilde{\mathscr{J}}_1(\bar{x}))_{ij}, \ \ \ \ \ \ \ \ \ \ \ \ \ \ \ j\neq i
  \end{array}
\right.
\end{align*} 
with 
 $(\widetilde{\mathscr{J}}_1(\bar{x}))_{ij}=(1-x_i(t))(\mu_{i}A_{ij}+\sum_{k=3}^{n} \mu_{ik}\sum_{i_2,\ldots, i_{k-1}\in N_{k}^{i}}A_{iji_2\ldots i_{k-1}}x_{i_2}(t)\ldots x_{i_{k-1}}(t)
 +\sum_{k=3}^{n} \mu_{ik}\\
 \sum_{i_1,i_3,\ldots, i_{k-1}\in N_{k}^{i}}A_{ii_1ji_3\ldots i_{k-1}}x_{i_1}(t)x_{i_3}(t)\ldots x_{i_{k-1}}(t)+\ldots
 +\sum_{k=3}^{n} \mu_{ik}\sum_{i_1,\ldots, i_{k-2}\in N_{k}^{i}}A_{ii_1\ldots i_{k-2}j}x_{i_1}(t)\ldots x_{i_{k-2}}(t)),$
  Observing that $\widetilde{J}_1(\bar{x})$ is a Metzler matrix. Besides, since $\bar{x}_i\geq\frac{n-2}{n-1}\geq\frac{k-2}{k-1}$ for $k=3,4,\ldots,n$, it is obvious that for all $i\in\mathcal{N}$,
\begin{align*}
(\widetilde{J}_1(\bar{x})\bar{x})_i=&-(\sum_{j\in\mathcal{N}}\mu_{i}A_{ij}\bar{x}_j)\bar{x}_i +\sum_{k=3}^{n}(k-2-(k-1)\bar{x}_i)\mu_{ik}\sum_{i_1,\ldots, i_{k-1}\in N_{k}^{i}}A_{ii_1\ldots i_{k-1}}x_{i_1}(t)\ldots x_{i_{k-1}}(t)\notag\\
\leq&-\mbox{min}_{i\in\mathcal{N}}(\sum_{j\in\mathcal{N}}\mu_{i}A_{ij}\bar{x}_j)\bar{x}_i
 \end{align*}
Then, there exists constant $\tilde{d}>0$, such that $(\widetilde{J}_1(\bar{x})\bar{x})_i\leq-\tilde{d}\bar{x}_i,$ which indicates that $\widetilde{J}_1(\bar{x})$ is a Hurwitz matrix. In other words, the following inequality holds: $s(\widetilde{J}_1(\bar{x}))<0.$

On the other hand, following that $\widetilde{J}(\bar{x})=I+h\widetilde{J}_1(\bar{x})$ is a nonnegative irreducible matrix and based on the Perron Frobenius Theorem, one gains that
there exists $\vartheta\gg0,$ such that
\begin{align*}
(I+h\widetilde{J}_1(\bar{x}))\vartheta=\rho(I+h\widetilde{J}_1(\bar{x}))\vartheta.
\end{align*}
Then,
\begin{align*}
h\widetilde{J}_1(\bar{x})\vartheta=\rho(h\widetilde{J}_1(\bar{x}))\vartheta=s(h\widetilde{J}_1(\bar{x}))\vartheta,
\end{align*}
which implies
\begin{align*}
hs(\widetilde{J}_1(\bar{x}))<0\Leftrightarrow\rho(I+h\widetilde{J}_1(\bar{x}))<1.
\end{align*}
Therefore, the endemic equilibrium point $\bar{x}$ is locally asymptotically stable.    {\hfill$\square$}

\subsection{ The derivation process of the error dynamics \eqref{error_general}}

\begin{align*}
y(t+1)&=x(t+1)-\bar{x}\\
      &=x(t)+h\left(-\mathcal{D}x(t)+(I-\mathsf{diag}(x(t))) (\mathcal{B}x(t)+\sum_{k=3}^{n}\mathcal{F}_kx^{k-1}(t))\right)-\bar{x}\\
      &=(I-h\mathcal{D})y(t)+h(I-\mathsf{diag}(\bar{x}))(\mathcal{B}y(t)+\sum_{k=3}^{n}\mathcal{F}_k\sum_{i=1}^{k-1}C_{k-1}^{i}\bar{x}^{k-1-i}y^i(t))\\
      &\ \ \ \ -h\mathsf{diag}(y(t))(\mathcal{B}y(t)+\mathcal{B}\bar{x}+\sum_{k=3}^{n}\mathcal{F}_k\bar{x}^{k-1}+\sum_{k=3}^{n}\mathcal{F}_k\sum_{i=1}^{k-1}C_{k-1}^{i}\bar{x}^{k-1-i}y^i(t))\\
&=\mathcal{G}_1y(t)+\mathcal{G}_2y^2(t)+\sum_{k=3}^{n}\sum_{i=3}^{k-1}\left(h(I-\mathsf{diag}(\bar{x}))\mathcal{F}_kC_{k-1}^{i}\bar{x}^{k-1-i}-h\widetilde{\mathcal{F}}_kC_{k-1}^{i-1}\bar{x}^{k-i}\right)y^{i}(t)-h\sum_{k=3}^{n}\widetilde{\mathcal{F}}_iy^{k}(t)\\     &=\mathcal{G}_1y(t)+\mathcal{G}_2y^2(t)+\sum_{i=3}^{n-1}\sum_{k=i+1}^{n}\left(h(I-\mathsf{diag}(\bar{x}))\mathcal{F}_kC_{k-1}^{i}\bar{x}^{k-1-i}-h\widetilde{\mathcal{F}}_kC_{k-1}^{i-1}\bar{x}^{k-i}\right)y^{i}(t)-h\sum_{i=3}^{n}\widetilde{\mathcal{F}}_iy^{i}(t)\\
&= \mathcal{G}_1 y(t)+\mathcal{G}_2 y^2(t)+\ldots +\mathcal{G}_n y^n(t).
\end{align*}
  
\end{document}